\documentstyle[emulateapj]{article}
\newcommand{\aliii}{\ifmmode{{\rm Al\,III}} \else Al\,III\fi}
\newcommand{\ciii}{\ifmmode{{\rm C\,III]}} \else C\,III]\fi}
\newcommand{\civ}{\ifmmode{{\rm C\,IV}} \else C\,IV\fi}
\newcommand{\hei}{\ifmmode{{\rm He\,I}} \else He\,I\fi}
\newcommand{\heii}{\ifmmode{{\rm He\,II}} \else He\,II\fi}
\newcommand{\lya}{\ifmmode{{\rm Ly}\alpha}\else Ly$\alpha$\fi}
\newcommand\lyb{\ifmmode {\rm Ly}\beta \else Ly$\beta$\fi}
\newcommand{\mgii}{\ifmmode{{\rm Mg\,II}} \else Mg\,I
I\fi}
\newcommand{\nv}{\ifmmode{{\rm N\,V}} \else N\,V\fi}
\newcommand{\feii}{\ifmmode{{\rm Fe\,II}} \else Fe\,II\fi}
\newcommand{\ovi}{\ifmmode{{\rm O\,VI}} \else O\,VI\fi}
\newcommand{\siiv}{\ifmmode{{\rm Si\,IV}} \else S\,IV\fi}
\newcommand\hb{\ifmmode {\rm H}\beta \else H$\beta$\fi}
\newcommand\hg{\ifmmode {\rm H}\gamma \else H$\gamma$\fi}
\newcommand\hd{\ifmmode {\rm H}\delta \else H$\delta$\fi}
\newcommand{\oiii}{\ifmmode{\rm [O\,III]} \else [O\,III]\fi}
\newcommand{\ew}{\ifmmode{W_{\lambda}} \else $W_{\lambda}$\fi}
\newcommand{\wciii}{\ifmmode{W_{\lambda}({\rm C\,III]})} \else $W_{\lambda}$(C\,III])\fi}
\newcommand{\wciv}{\ifmmode{W_{\lambda}({\rm C\,IV})} \else $W_{\lambda}$(C\,IV)\fi}
\newcommand{\wfeii}{\ifmmode{W_{\lambda}({\rm Fe\,II})} \else $W_{\lambda}$(Fe\,II)\fi}
\newcommand{\wheii}{\ifmmode{W_{\lambda}({\rm He\,II})} \else $W_{\lambda}$(He\,II)\fi}
\newcommand{\wlya}{\ifmmode{W_{\lambda}({\rm Ly}\alpha)}\else $W_{\lambda}$(Ly$\alpha$)\fi}
\newcommand\wlyb{\ifmmode{ W_{\lambda}({\rm Ly}\beta )} \else $W_{\lambda}$(Ly$\beta$)\fi}
\newcommand{\wmgii}{\ifmmode{W_{\lambda}({\rm Mg\,II})} \else $W_{\lambda}$(Mg\,II)\fi}
\newcommand{\wovi}{\ifmmode{W_{\lambda}({\rm O\,VI})} \else $W_{\lambda}$(O\,VI)\fi}
\newcommand{\wsiiv}{\ifmmode{W_{\lambda}({\rm Si\,IV})} \else $W_{\lambda}$(Si\,IV)\fi}
\newcommand\whb{\ifmmode{ W_{\lambda}({\rm H}\beta )} \else $W_{\lambda}$(H$\beta$)\fi}
\newcommand\whg{\ifmmode{ W_{\lambda}({\rm H}\gamma )} \else $W_{\lambda}$(H$\gamma$)\fi}
\newcommand\whd{\ifmmode{ W_{\lambda}({\rm H}\delta )} \else $W_{\lambda}$(H$\delta$)\fi}
\newcommand{\woii}{\ifmmode{{W_{\lambda}(\rm [O\,II])}} \else $W_{\lambda}$([O\,II])\fi}
\newcommand{\woiii}{\ifmmode{{W_{\lambda}(\rm [O\,III])}} \else $W_{\lambda}$([O\,III])\fi}\newcommand{\kms}{\ifmmode~{\rm km~s}^{-1}\else ~km~s$^{-1}~$\fi}

\lefthead{Kuraszkiewicz, Green, Crenshaw, Dunn, Forster, Vestergaard, Aldcroft}
\righthead{Emission Line Properties of AGN from a post-COSTAR FOS HST
Spectral Atlas}

\begin{document}

\title{Emission Line Properties of AGN from a post-COSTAR FOS 
HST Spectral Atlas}

\author{Joanna K. Kuraszkiewicz, Paul J. Green}
\affil{Harvard-Smithsonian Center for Astrophysics, 60 Garden St.,
Cambridge, MA 02138} 
\affil{email: {\em jkuraszkiewicz@cfa.harvard.edu, pgreen@cfa.harvard.edu}}

\author{D. Michael Crenshaw, Jay Dunn}
\affil{Department of Physics and Astronomy, Georgia State University,
Astronomy Offices, One Park Place South SE, Suite 700, Atlanta, GA
30303}
\affil{email: {\em crenshaw@chara.gsu.edu,dunn@chara.gsu.edu}}

\author{Karl Forster}
\affil{California Institute of Technology, 1200 E. California Blvd., 
MC 405-47, Pasadena, CA 91125}
\affil{email: {\em krl@srl.caltech.edu}}

\author{Marianne Vestergaard}
\affil{The Ohio State University, Columbus, 140 West 18th Avenue,
Columbus, OH 43210} 
\affil{email: {\em vester@astronomy.ohio-state.edu}}

\author{Tom L. Aldcroft}
\affil{Harvard-Smithsonian Center for Astrophysics, 60 Garden St.,
Cambridge, MA 02138} 
\affil{email: {\em taldcroft@cfa.harvard.edu}}

\begin{abstract}

This paper joins a series compiling consistent emission line
measurements of large AGN spectral databases, useful for reliable
statistical studies of emission line properties. It is preceded by
emission line measurements of 993 spectra from the Large Bright Quasar
Survey (Forster et al. 2001) and 174 spectra of AGN obtained from the
Faint Object Spectrograph (FOS) on HST prior to the installation of
COSTAR (Kuraszkiewicz et al. 2002).  This time we concentrate on 220
spectra obtained with the FOS after the installation of COSTAR,
completing the emission line analysis of all FOS archival spectra.  We
use the same automated technique as in previous papers, which accounts
for Galactic extinction, models blended optical and UV iron emission,
includes Galactic and intrinsic absorption lines and models emission
lines using multiple Gaussians.  We present UV and optical emission
line parameters (equivalent widths, fluxes, FWHM, line positions) for
a large number (28) of emission lines including upper limits for
undetected lines. Further scientific analyses will be presented in
subsequent papers.
\end{abstract}


\section{Introduction}

It is broadly acknowledged that the quasar central engine (presumably
a massive black hole with an accretion disk) photoionizes gas lying
farther out. This gas emits broad permitted emission lines that are
distinctive of quasar spectra.  At first glance, quasar spectra look
quite similar; this may be the result of simple averaging.  Baldwin et
al. (1995) showed that although the broad line region (BLR) consists
of clouds with a wide range of properties (gas density, ionization
flux and column density), the bulk of emission line flux is most
likely produced in the gas clouds with the optimum parameters for
efficient emission in that line.

A closer look at the quasar spectra, however, reveals that the spectra
differ in detail and intriguingly, behave in a correlated manner.  For
example it was found that AGN that show strong optical iron emission
(\ion{Fe}{2}\,$\lambda 4570$) have weaker [\ion{O}{3}]\,$\lambda
5007$, and narrower, blue-asymmetric H$\beta$ lines.  This set of
correlations was found to be the primary eigenvector of the
emission line correlation matrix of PG quasars studied by Boroson \&
Green (1992).  This eigenvector~1 was later found to correlate with UV
properties such as: CIV shift/asymmetry (Marziani et al. 1996) and
\ion{Si}{3}]/\ion{C}{3}] ratio, \ion{C}{4} and \ion{N}{5} strength
(Wills et al. 1999, Shang et al. 2003). Since eigenvector~1 was found to 
correlate significantly with X-ray properties (Laor et al. 1997,
Brandt \& Boller 1998) which are determined in the vicinity of the central
black hole, it was suggested that differences in emission line
properties revealed by eigenvector~1 are caused by differing central engine
parameters (e.g. $L/L_{Edd}$, accretion rate, orientation and/or black
hole spin). It was found that eigenvector~1 together with
eigenvector~2 provide a parameter-space in which all major classes of
broad-line sources can be discriminated, constituting a possible ``H-R
diagram'' for quasars (Sulentic et al. 2000; Boroson 2002).

Another famous correlation involving quasar spectra is the
anticorrelation between the equivalent width of the broad emission
lines and the UV luminosity called the Baldwin effect (Baldwin
1977). The appeal of this correlation was soon realized, since the
luminosity of a distant quasar could potentially be estimated from the
emission line equivalent widths, providing a standard candle in
measuring cosmological distances. In reality the scatter of the
Baldwin effect is too large to give meaningful results, and studies
have concentrated on understanding and reducing this scatter (Shang et
al. 2003, Dietrich et al. (2002).  Conflicting results have also
emerged, where radio-loud samples and samples with a wide range of
luminosities show a stronger effect (e.g. Baldwin et al. 1978, Wampler
et al. 1984, Kinney, Rivolo, \& Koratkar 1990, Wang et al. 1998),
while radio-quiet samples and samples with a small luminosity range
show weaker or no effect (e.g. Steidel \& Sargent 1991, Wilkes et
al. 1999). A number of explanations have been introduced to explain
the Baldwin effect. It can be either due to geometry as in Netzer,
Laor \& Gondhalekar (1992), where the inclination of the disk changes
the apparent luminosity, or due to changes in spectral energy
distribution with luminosity, where more luminous objects have softer
ionizing continuum (Zheng \& Malkan 1993, Green 1998) or due to a
decrease of covering factor of the broad emission line clouds with
luminosity (Wu, Boggess, \& Gull 1983).  There have also been claims
that the Baldwin effect is affected by evolution (Green, Forster, \&
Kuraszkewicz 2001) or may be due to selection effects (continuum
beaming, biases in selection techniques - see Sulentic et al. 2000,
Yuan, Siebert, \& Brinkmann 1998).

Despite a vigorous study of emission line properties of AGN in the
last 30 years, which resulted in few thousand published articles,
questions about the structure and kinematics of the BLR and their
relationship to the central engine (accretion mechanism, origin of the
fuel, etc.)  have not been answered. Nor is it clear how the BLR
relates to the other components seen in AGN spectra: broad and narrow
absorption lines, X-ray warm absorbers, high ionization emission lines
and scattering regions. Despite attempts to unite these components
(Elvis 2000, Laor \& Brandt 2002, Ganguly et al. 2001, Murray \& Chiang
1995) definitive tests have been elusive.  Progress has been
hampered by lack of large datasets with uniform and reliable
measurements of emission lines that would consistently measure the
continuum, and account for blended iron emission which heavily
contaminates emission lines such as: H$\beta$, \ion{Mg}{2}, and
\ion{C}{3}] and forms a pseudo-continuum complicating the measurements
of the broadband continuum, the 
weaker lines and the wings of strong emission lines (Wills et
al. 1985, Boroson \& Green 1992, Vestergaard \& Wilkes 2001). Most
studies have concentrated either on large non-uniform samples where
emission line measurements have been compiled from literature (Zheng
\& Malkan 1993, Zamorani et al. 1992, Corbin \& Boroson 1996, Dietrich
et al. 2002) or small samples with uniform measurements (Boroson \&
Green 1992, Wills et al. 1999, Wilkes et al. 1999).

We have therefore undertaken a major study of AGN emission lines,
where our largely automated procedure accounts for Galactic
extinction, models blended optical and UV iron emission, includes
Galactic and intrinsic absorption lines, and models emission lines
using multiple Gaussians. Using the same modeling procedure we
have previously analyzed and published measurements of emission lines
of two large datasets. The first, 993 spectra from the Large
Bright Quasar Survey has been presented by Forster et al. (2001;
hereafter Paper~I) together with detailed description of our
analysis methods.  The second includes 174 FOS/HST spectra
obtained before the installation of COSTAR and was presented in
Kuraszkiewicz et al. (2002; hereafter Paper~II).  In the current paper we
present the measurements of emission lines and plots of spectral fits
of the remaining 220 FOS/HST spectra that were observed after the
installation of COSTAR, completing the analysis of all archival
FOS/HST spectra. Statistical comparison of the emission-line
parameters and continuum parameters of these large samples will
hopefully bring us closer to building an accurate model of emission
line regions and their dependence on the central engine.

\section{The Post-COSTAR FOS AGN Sample}

The sample was assembled by cross-correlating the Veron-Cetty and
Veron (1996) catalog of AGN with the MAST (Multimission Archive at
Space Telescope) holdings. BL Lac objects were ignored, as their
spectra show no emission lines. Starburst galaxies and broad
absorption line (BAL) quasars (where emission lines are heavily
disrupted by absorption features) were not included. We chose all
available (UV and optical) spectrophotometric archival data that have
been observed with the Faint Object Spectrograph (FOS, Keyes et
al. 1995 and references therein) on HST after the installation of
COSTAR (i.e., after December 1993).  FOS spectra obtained prior to
December 1993 have been analyzed by us in Paper~II. We include all
spectra taken with the high resolution gratings (G130H, G190H, G270H,
G400H, G570H, G780H; spectral resolution $\lambda/\Delta\lambda
\sim 1300$ ).  Low resolution (G160L, G650L; spectral resolution
$\lambda/\Delta\lambda \sim 250$) gratings were also included when
high resolution gratings were not available in the matching wavelength
range.  Spectra obtained with the prism were excluded as their
extremely low resolution precludes any reasonable emission line
measurements.
We analyzed only spectra with a mean signal-to-noise (S/N) per
resolution element $\ge 5$.

The FOS spectra were uniformly calibrated to account for temporal,
wavelength-- and aperture--dependent variations that are seen in the
instrumental response.  We use the most recent version of the FOS
calibration pipeline with the ST-ECF POA version of {\it calfos}.  This
pipeline provides an improved correction to the zero point offsets in
the BLUE high resolution spectra, removes hot pixel/hot diode regions
from individual exposures and calibrates spectra to the 4.3''
aperture.

We interpolated all of the spectra to a linear wavelength scale,
retaining the original approximate wavelength intervals (in Angstroms
per bin), and, for each object, we averaged all of the spectra
obtained at a particular wavelength setting if the flux did not differ
by more than 20\%. If the difference in flux was larger, the spectra
were analyzed separately. To obtain a reliable continuum fit for each
object, we combined spectra obtained at different wavelength settings
and observed at different times if the flux levels did not differ by
more than 20\% in the overlap region.  High resolution gratings
(G130H, G190H, G270H, G400H, G570H, G780H) were merged separately from
the low resolution gratings (G160L, G650L). In both cases the longer
wavelength spectrum was scaled to match the shorter wavelength
spectrum and the spectra were then spliced at wavelengths in continuum
regions away from emission or absorption lines by retaining as much of
the higher S/N spectrum as possible.

At this point the sample consisted of 327 spectra. Spectra which
showed no emission lines, due to a too low S/N ($<$5) in the line
regions, or a redshift
that placed strong emission lines outside the spectrum's wavelength
range (mostly chosen for studies of the Ly$\alpha$ forest) were then
removed.  The final sample consists of 220 spectra of the 
180 AGN listed in Table~1.  In the first column the coordinate designation
based on the equinox J2000 position is given, followed by the AGN name
(column [2]), AGN type and redshift (from the NASA/IPAC Extragalactic
Database) and Galactic $N_{H}$ in units of $10^{20}$cm$^{-2}$
(columns [3]-[5]). The values of $N_{H}$ are in general taken from the
Bell Laboratory \ion{H}{1} survey (Stark et al. 1992). In a few cases
for which $N_H$ had been specifically measured, we quote the values
from the literature (Lockman \& Savage 1995, Elvis et al. 1989); for
objects with declination $> 40^o$, $N_H$ is from Heiles \& Cleary
(1979). The last column of Table~1 gives the list of spectra that were
analyzed for each object. The name of the spectrum consists of the
coordinate designation from column (1), followed by a two letter
designation: ``o'' indicates a post-COSTAR spectrum (in
Paper~II pre-COSTAR spectra were designated with ``r''); a second
letter (a to z) indicates whether the AGN in question has more
than one spectrum available.  A capital letter indicates a spectrum of
a lensed component as e.g. in 1001+5553oA and 1001+5553oB. In Table~2
we show a detailed list of FOS gratings, and datasets with exposure
times that were used to compile spectra listed in Table~1 (see 
the ApJ Web site\footnote{http://www.journals.uchicago.edu/ApJ}
for full version of Table~2).

\section{Analysis of Spectra}

\subsection{Continuum and blended iron fitting}

Since our goal was to assemble a uniform database of emission line
measurements, we have analyzed our post-COSTAR spectra following the
same fitting procedures as those used in the LBQS and pre-COSTAR/FOS
spectral analysis (for details see Papers~I and II).  We used the
modeling software {\em Sherpa}\footnote
{http://cxc.harvard.edu/sherpa/index.html} (Freeman, Doe \&
Siemiginowska 2001) developed for the Chandra mission, where the model
parameters were determined from a minimization of the $\chi^2$
statistic with modified calculation of uncertainties in each bin
(Gehrels 1986) and using the Powell optimization method for continuum,
iron emission and first emission line fits and the Levenberg-Marquardt
optimization method in the final emission line fits (see below). First
we fit a reddened power-law continuum\footnote{We use the the
reddening curves of Cardelli, Clayton \& Mathis 1989 to account for
Galactic extinction; see Paper~I for details} to regions of the
spectrum redwards of Ly$\alpha$ and away from strong emission lines
and blended iron emission. We use the same continuum windows as in the
analysis of pre-COSTAR continuum spectra (see Table~2 in Paper~II),
with the addition of a new window redwards of H$\alpha$ at
6990--7020\AA\ rest frame. Most of the post-COSTAR spectra were fitted
by a single power law. However in 21 spectra that covered a large
wavelength range, two power laws were introduced: one (UV) extending
at $\lambda_{rest}<4200$\AA\ and another (optical) at
$\lambda_{rest}>4200$\AA, both 
normalized at $\lambda=4200$\AA. In Table~3 we present the slopes of
the dereddened UV and optical continua (column [2] and [5]
respectively) with the normalization of the continuum in units of
10$^{-14}$~erg~cm$^{-2}$s$^{-1}$\AA$^{-1}$ (column [3]) at the
observed wavelength $\lambda_{norm}$ (column [4]). The slopes and
normalizations are quoted with 2$\sigma$ errors. For spectra with only
one continuum window present, a constant slope of $\Gamma=1$ is quoted
without errors. This value was adopted since the mean slope of the pre
and post-COSTAR FOS sample is 0.97$\pm$0.09 (see the ApJ Web
site\footnote{http://www.journals.uchicago.edu/ApJ} for full
version of Table~3).

The next step in our fitting procedure was to model the blended iron
emission lines.  In the UV we used the Vestergaard \& Wilkes (2001)
iron template covering rest frame wavelengths between 1250--3100\AA,
while in the optical we used the Boroson \& Green (1992) template
covering 4250--7000\AA. First a crude estimate of the template's flux
normalization was obtained by fitting the 2000\kms\, FWHM template to
regions where iron emission is known to be strongest (see column~[2]
in Table~2 of Paper~II). Then the FWHM of iron emission was estimated
by comparing the spectrum with a grid of templates with FWHM between
900 and 10,000\kms in steps of 250\kms. This was followed by a fit of
both the FWHM and flux normalization at the iron fitting windows,
followed by two iterations of the continuum and iron fits (refer to
Paper~I for more details).  At this point the continuum and iron fits
results were inspected and adjustments were made to spectra not fitted
successfully (5\% of spectra needed adjustments of the continuum fit
and 3 spectra needed adjustment of iron fits).

\subsection{Emission and absorption line fitting}

The emission lines were generally fitted with one Gaussian. However
since most (95\%) FOS spectra have high S/N, the strong emission lines
(Ly$\alpha$, C\,IV, C\,III], Mg\,II, H$\beta$, H$\alpha$) were fitted
using two components: the very broad line region (VBLR) component and
the intermediate line region (ILR; see Brotherton et al. 1994) here
referred to as the broad and narrow components respectively. We use
exactly the same emission line inventory as in the pre-COSTAR spectra
(see Table~3 in Paper~II).

As a first stage, the FWHM and peak amplitude of the Gaussians are
modeled while keeping the position of the emission line fixed at the
expected wavelength (calculated from redshift). Then the position of
the line is freed and modeled together with the FWHM and peak
amplitude using Powell optimization. In the next step all Gaussian
parameters are refitted, this time using distinct high and low {\it
sigma rejection} criteria.  We found that $\sigma=3$ for low rejection
omits most of the absorption lines superimposed on the emission lines,
while $\sigma=7$ for high rejection bypasses most spikes not
associated with the emission line (e.g. geocoronal Ly$\alpha$, cosmic
rays, etc.). At this stage we use the Levenberg-Marquardt optimization
method, which is faster than the Powell method, but only works well if
the statistical surface is well-behaved (after two runs of the
emission line parameter fitting with the Powell method, this was
certainly the case).  It is nearly impossible to design a fully
automated procedure that can deal with the wide range of spectral
shapes that AGN show, so at this point the fits were inspected and
adjustments were made to spectra where necessary. About 5\% of spectra
needed adjustments at least in one emission line fit.

For each spectrum, the continuum, iron and emission line model
obtained in the {\em Sherpa} fitting was next used as an input
``continuum'' to the FINDSL routine (Aldcroft 1993), which identifies
narrow absorption lines and fits them with Gaussian profiles.  We set
the routine to find absorption lines away from the Ly$\alpha$ forest
region (blueward of $\lambda_{rest} =$1065\AA) and outside the Balmer
continuum region (3360--3960\AA), where the global power law continuum
may not fit the spectra well.  The minimum significance level for
identification of absorption lines was set to 4.5$\sigma$ (see Paper~I
for more details). We detect and fit absorption lines with \ew$\ge
0.3$\AA. The absorption line parameters were then used in the next
iterative modeling step where the position, peak amplitude, and FWHM
of the absorption line were modeled simultaneously by the {\em Sherpa}
program, followed by another iteration of the emission line
fitting. After this stage the results were inspected and spectra
refitted if the automated procedure did not perform well. An example
of a full spectral fitting is shown in Fig.1. The top panel shows the
reddened power-law continuum fit redwards of Ly$\alpha$ to the
observed spectrum, followed below by panels showing blended iron and
emission line modeling of Ly$\alpha$, C\,IV, C\,III], and Mg\,II.
Since the whole post-COSTAR FOS Spectral Atlas includes 220 spectra,
we present similar plots of other spectral fits only on our Web
site.\footnote{See http://hea-www.harvard.edu/\~{}pgreen/HRCULES.html}

\subsection{Error analysis}

The error analysis follows the procedure from Paper~I (see Section~3.5
of that paper for details), in which the 2$\sigma$ errors for each
emission line parameter were determined from the $\chi^2$ confidence
interval bounds ($\Delta\chi^2$=4.0) using the {\it uncertainty}
procedure in {\em Sherpa}. The upper limits of equivalent widths were
determined by fixing the line position at the expected wavelength, the
FWHM at the value of the median FWHM found for that line in the LBQS
sample (see column [3], Table~3 of Paper~II) and by setting the
amplitude of the line to the 2$\sigma$ positive error.

\section{Emission line measurements}

In Table~4 we present the rest frame emission line measurements for
one example object NGC~3516 (spectrum 1106+7234oe).  Due to its large
size the full table for the post-COSTAR FOS Spectral Atlas is
available only in electronic from the ApJ Web
site\footnote{http://www.journals.uchicago.edu/ApJ} and at our
Web site$^5$.  The format of Table~4 is exactly the same as the format
of the electronic tables of emission line measurement presented for
the LBQS and pre-COSTAR FOS samples, making it simple to analyze the
LBQS and FOS samples together. In the full Table~4, each spectrum is
represented by 43 rows, one for each possible emission line
measurement. In the first column the name of the spectrum is given,
followed by the object's redshift (column [2]), followed by
information on the emission line parameters: name of the emission line
(column [3]), FWHM in \kms (columns [4] to [6] showing the value and
$\pm$2$\sigma$ errors), the offset of the peak of the Gaussian
emission line model (all lines except iron) in \kms from the expected
position based on the tabulated redshift (columns [7] to [9] value,
$\pm$2$\sigma$ errors), the rest frame equivalent width of the
emission line in \AA\ (columns [10]-[12]) and the observed frame flux
in units of 10$^{-14}$erg\,cm$^{-2}$\,s$^{-1}$ (columns [13]-[15]).
Errors quoted for flux and \ew\ are based on the uncertainties in the
amplitude and FWHM of the Gaussian model and do not include an error
from an uncertainty in the underlying continuum flux level, which we
estimate to be about 10\%. For emission lines where only an upper
limit on flux and \ew\ is available, no values for the peak offset are
quoted as the position of the line was fixed at the line's expected
wavelength. Also, the FWHM value in this case was set to the median
value for the LBQS sample (see Table 3 in Paper~II) with no associated
errors. Finally, the last column (16) in the full table gives the
number of narrow absorption features used in the emission line
modeling. Our Gaussian decomposition is not necessarily unique and may
be sensitive to slight shifts in continuum placement.  While the total
flux and equivalent width are easily derived by summing values
provided for individual Gaussian components, no simple combination
yields a FWHM representative of the entire emission line. We therefore
list in Table 4b (electronic version only) the total line FWHM (with
$\pm$2$\sigma$ errors) of those lines that have been modeled using two
Gaussians. These are: Ly$\alpha$, \ion{C}{4}, \ion{C}{3}],
\ion{Mg}{2}, H$\beta$, and H$\alpha$ where the width of the line was
measured at half peak of the dereddened emission line model after
excluding iron emission, absorption lines, and weaker emission lines
(e.g. in the Lya region we exclude NV line, in H$\alpha$ region [NII]
and [SII]).

\section{Statistics and comparison with the post-COSTAR sample}

The statistical properties of the rest frame \ew\ and FWHM
distributions of the emission line measurements of the post-COSTAR
spectra are presented in Table~5.  The numbers quoted were obtained by
excluding off-nuclear spectra (e.g. the 10 different NLR knots of
Mrk~78).  To avoid excessive weight given to a single object, in cases
of multiple spectra we tally only measurements from the spectrum with
the highest S/N and resolution.  In total 1607 emission lines have
been modeled among which 
97 are upper limits.  In Table~(5) the name of the emission line is
given in column (1), followed by the total number of emission lines
modeled (column~[2]) and the number of upper limits (column [3]). The
mean, standard deviation and median of the \ew\ and FWHM for the
detected lines are presented in columns (4)-(6) and (10)-(12)
respectively. When upper limits in \ew\ were present we used the
non-parametric survival analysis technique and a Kaplan-Meier
estimator to reconstruct the true \ew\ distribution and to calculate
the means and medians in columns (7)-(8) (for reference see Isobe,
Feigelson \& Nelson 1986 and Lavalley, Isobe \& Feigelson 1992).

Since the strong emission lines such as: Ly$\alpha$, \ion{C}{4},
\ion{C}{3}], \ion{Mg}{2}, H$\beta$, H$\alpha$ were fitted using either
two (broad and narrow) components or one (single) component, we
calculated the \ew\ and FWHM means and medians for these components
separately.  The mean and median \ew\ for the whole line (indicated as
the ``sum'' in Table~5) was calculated as either the sum of the broad and
narrow components or the single component alone.

Both the pre- and post-COSTAR FOS samples are heterogeneous, and
represent neither complete nor uniform selection.  Nevertheless, as a
check on our methods and on the consistency between these samples we
compare the statistical properties of the \ew\ and FWHM of the
pre-COSTAR sample analyzed in Paper~II and the post-COSTAR sample
presented here.  Overall, the means and medians for the UV lines agree
within the errors. We did not, however, attempt to compare the \ew\ of
post-COSTAR single components of the strong emission lines or optical
lines redwards of \ion{Ne}{5} with the pre-COSTAR sample measurements,
since the number of available \ew\ in both or either samples is too
small for meaningful analysis.
 
Histograms of \ew\ and FWHM of the emission lines bluewards of \ion{Mg}{2}
are presented in Figure~2. The first and third rows represent the \ew\
distributions, while the second and fourth rows give the FWHM
distributions. In all panels, solid lines represent distributions for
detections, while the dotted lines show the estimated \ew\
distributions from the Kaplan-Meier estimator if upper limits are
present. In the panels which show the sum of Ly$\alpha$, C\,IV,
C\,III] and Mg\,II distributions, the shaded histograms represent
results from single Gaussian component fits.

The luminosity and redshift range of the post-COSTAR sample is
comparable to the pre-COSTAR sample analyzed in Paper~II (see
Fig.3). However the post-COSTAR sample shows a larger number of low
luminosity AGN such as Seyferts, LINERs and NLS1s. Objects with
log\,$L(2500$\AA)$< 30$ comprise of $\sim$30\% of the post-COSTAR sample
and only 15\% of the pre-COSTAR sample. In Figure~4 we show the
distributions of log\,$L(2500$\AA$)$ for both samples. The two-tailed
Kolmogorov-Smirnow test gave a 99.9\% probability that these
distributions are different.

\section{Conclusions}

We have presented the emission line measurements of a sample of AGN
which has been observed by the FOS/HST after the installation of
COSTAR. Our sample includes 180 objects and 220 spectra, that have
been modeled using an automated technique which fits multiple
Gaussians to the emission lines, taking into account Galactic
reddening, blended iron emission, and Galactic and intrinsic
absorption lines. In this paper we present uniform measurements of
1607 emission lines including equivalent widths, FWHM and shifts from
the line's expected position and calculate upper limits for weak
lines. We also present the underlying continuum parameters (slopes and
normalization). This is the third paper in a series of papers aimed at
uniformly measuring emission line properties in large AGN samples. It
has been preceded by a presentation of emission line properties in
$\sim$1000 optical spectra from the Large Bright Quasar Survey
(Paper~I) and $\sim$200 UV spectra observed by FOS/HST in the
pre-COSTAR era (Paper~II). All 1387 spectral fits and tabulated
results are available at our Web
site.\footnote{http://hea-www.harvard.edu/\~{}pgreen/HRCULES.html}
Such large uniformly measured databases will hopefully bring us closer
to a better understanding of the origin of the line emitting regions
and their relationship to the central engine.

\acknowledgements

PJG and JK gratefully acknowledge support provided by NASA through
grant NAG5-6410 (LTSA). PJG and TA acknowledges support through NASA
contract NAS8-39073 (CXC).  MV acknowledges financial support for
Proposal number AR-09549, provided by NASA through a grant from the
Space Telescope Science Institute, which is operated by the
Association of Universities for Research in Astronomy, Incorporated,
under NASA contract NAS5-26555.  We are grateful to Todd Boroson for
providing the FeII optical template.  This research was made based on
observations made with the NASA/ESA Hubble Space Telescope, obtained
from the data archive at the Space Telescope Science Institute and
using the Multimission Archive at the Space Telescope Science
Institute (MAST). STScI is operated by the Association of Universities
for Research in Astronomy, Inc., under NASA contract
NAS5-26555. Support for MAST for non-HST data is provided by the NASA
Office of Space Science via grant NAG5-7584 and by other grants and
contracts.  This research has also made use of the NASA/IPAC
Extragalactic Database (NED) which is operated by the Jet Propulsion
Laboratory, California Institute of Technology, under contract with
the National Aeronautics and Space Administration.

\appendix

Notes on individual objects:

0039-5117oa - two power law continua were fitted to this spectrum,
which were joined at a non-standard wavelength of 3000\AA\ observed
frame, for a  better continuum fit.  

0238+1636oa - this BL Lac object is included in our sample as it shows
weak emission lines.  

0241-0815oa - spectrum spans a large wavelength range from 2200\AA\ to
6800\AA, so the power law continuum does not fit the spectrum well,
especially at H$\alpha$ wavelengths.

0320-1926oa, 1337+2423oa, 1959+4044oa- short spectra with only one standard
continuum window, for a better continuum fit we added a non-standard
continuum window at red side of MgII.

0742+6510oa-oj - spectra of 10 different NLR knots in Mrk~78 (a
Seyfert~2) showing interaction of the NLR gas with the ISM.

1048-2509oa - very weak continuum. 

1223+1545oa - very weak continuum. 

1252+2913oa - used a nonstandard continuum window at wavelengths 2100 to
2130\AA\ observed frame. 

1719+4858oa - missing spectrum at CIV wavelengths.

1842+7946oa - missing spectrum at CIV wavelengths.

1902+3159oa - added additional window blue of Ly$\alpha$ for better
continuum fit. 

1927+7358oa - missing spectrum at Ly$\alpha$ wavelengths.


\clearpage

\begin{figure}
\epsscale{.90}
\plotone{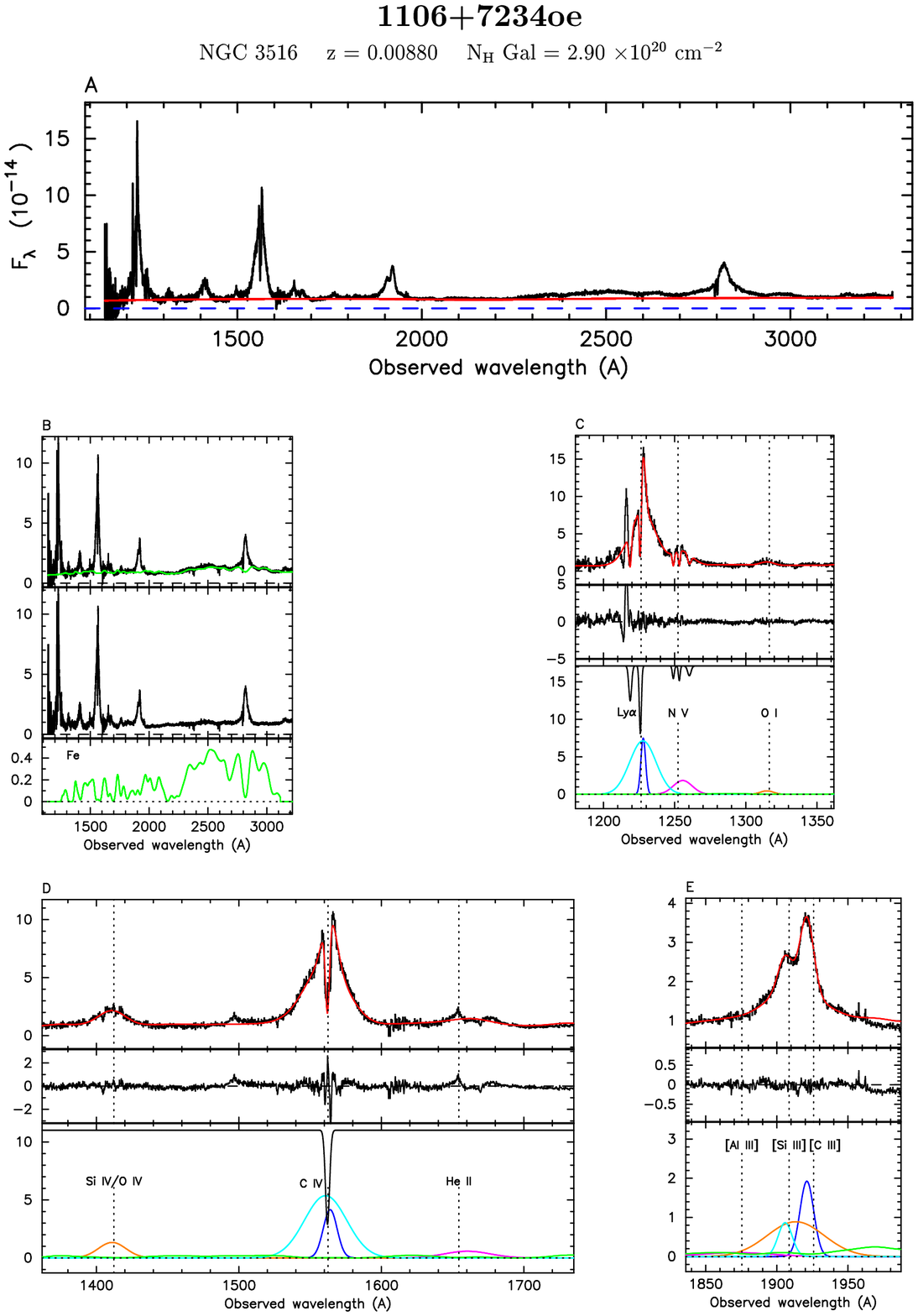}
\caption{\footnotesize {Example of spectral modeling of NGC~3516.  Panel A
shows the reddened power law continuum model fitted redwards of
Ly$\alpha$ plotted over the observed spectrum. Below in panel B we
show iron modeling, divided into 3 frames: top - showing the
continuum+iron model plotted over the overall spectrum, middle showing
the iron subtracted spectrum, and bottom showing the fitted iron
template. Panels C,D, and E show modeling of the Ly$\alpha$, C\,IV and
C\,III] emission line regions respectively.  Each panel for each
emission line region (C,D,E) is divided into 3 frames: top frame
shows the total best-fit model plotted over the relevant region of
each spectrum, middle frame the residuals, and bottom frame the
individual Gaussian components. Strong emission lines such as
Ly$\alpha$, C\,IV, and C\,III] are modeled with two components: narrow
and broad, while other emission lines are modeled using one Gaussian.
The absorption lines that overlap each emission line are plotted at
the top of the bottom frame. The dashed vertical lines in the emission
line panels are drawn at the expected emission line position
calculated using the redshift quoted at the top of the figure.  Flux
units are $10^{-14}$ erg~cm$^{-2}$s$^{-1}$ \AA$^{-1}$, wavelength
units are in \AA~ and are observed frame values.}}
\end{figure}

\clearpage
\voffset=0cm

\begin{figure}
\epsscale{1.0}
\plotone{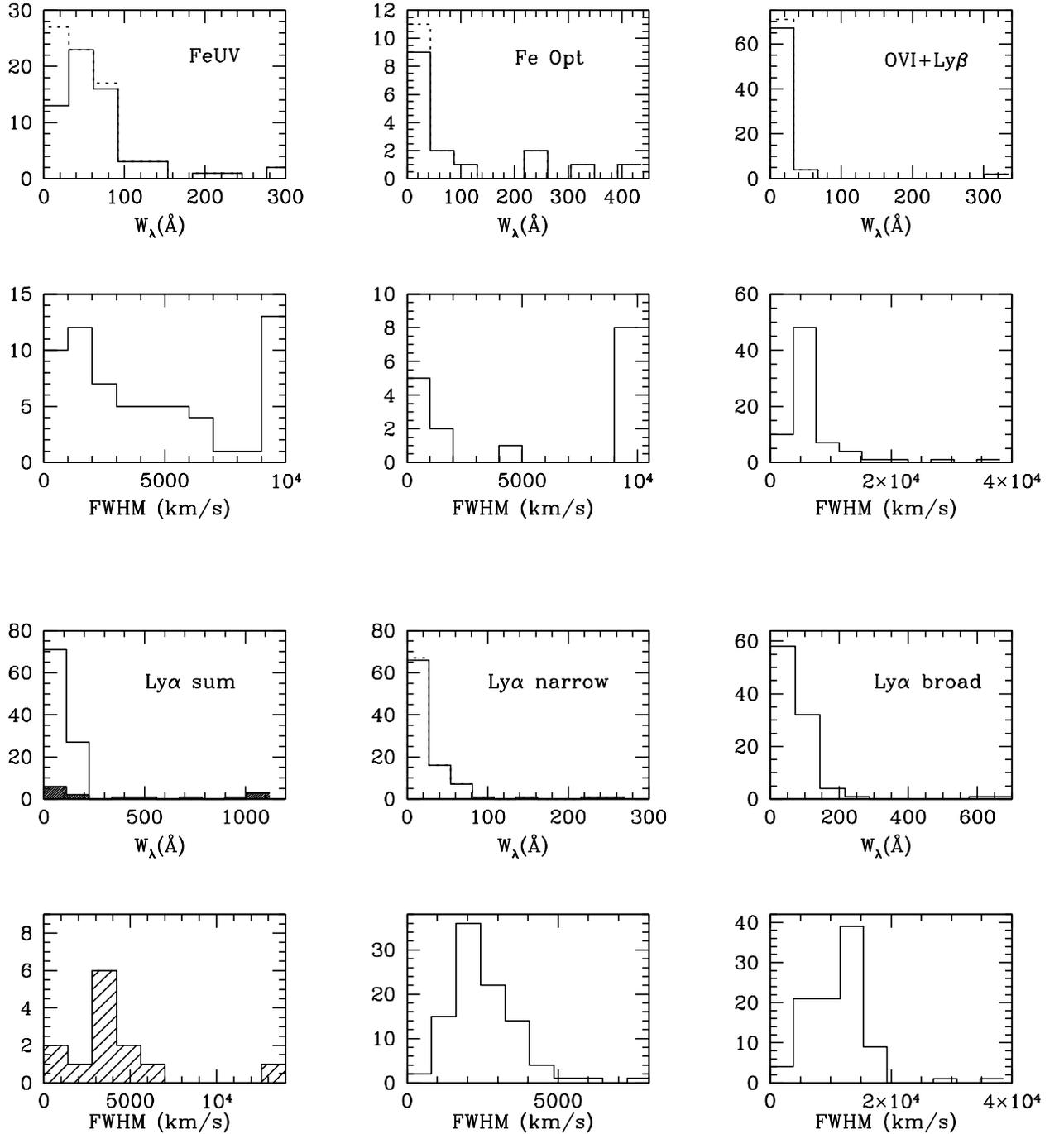}
\caption{The distribution of rest frame \ew\ (top and third row) and
FWHM (second from top and bottom row) of the emission line properties
of AGN in the post-COSTAR FOS sample.  When upper limits in \ew\ are
present we show the estimated Kaplan-Meier distributions with a dashed
line. Shaded areas represent distributions for the single Gaussian
components of the strong emission lines: Ly$\alpha$, \ion{C}{4},
\ion{C}{3}] and \ion{Mg}{2}.}
\end{figure}

\clearpage

\begin{figure}
\plotone{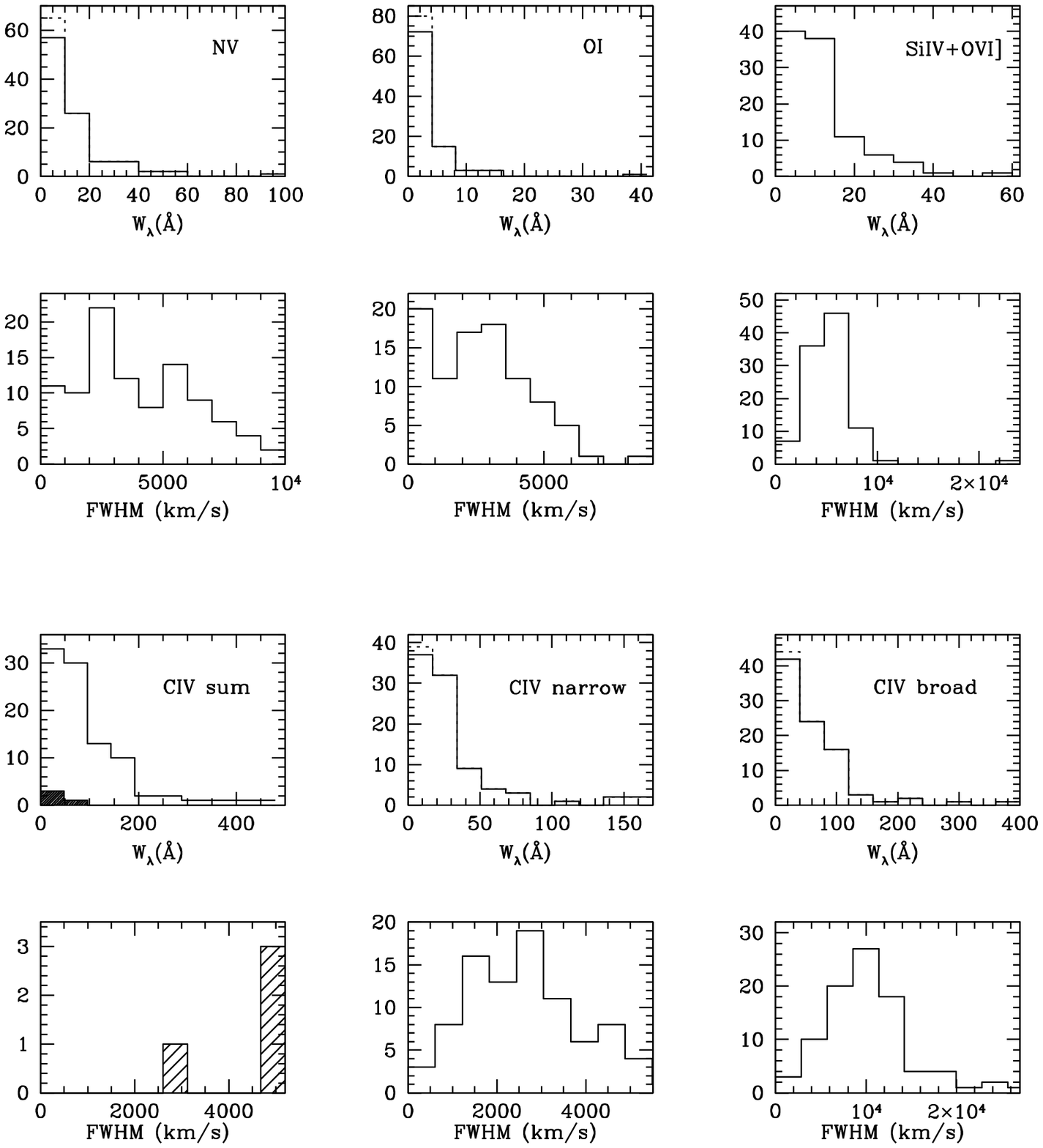}
\setcounter{figure}{1}
\caption{{\it Continued}}
\end{figure}

\clearpage

\begin{figure}
\plotone{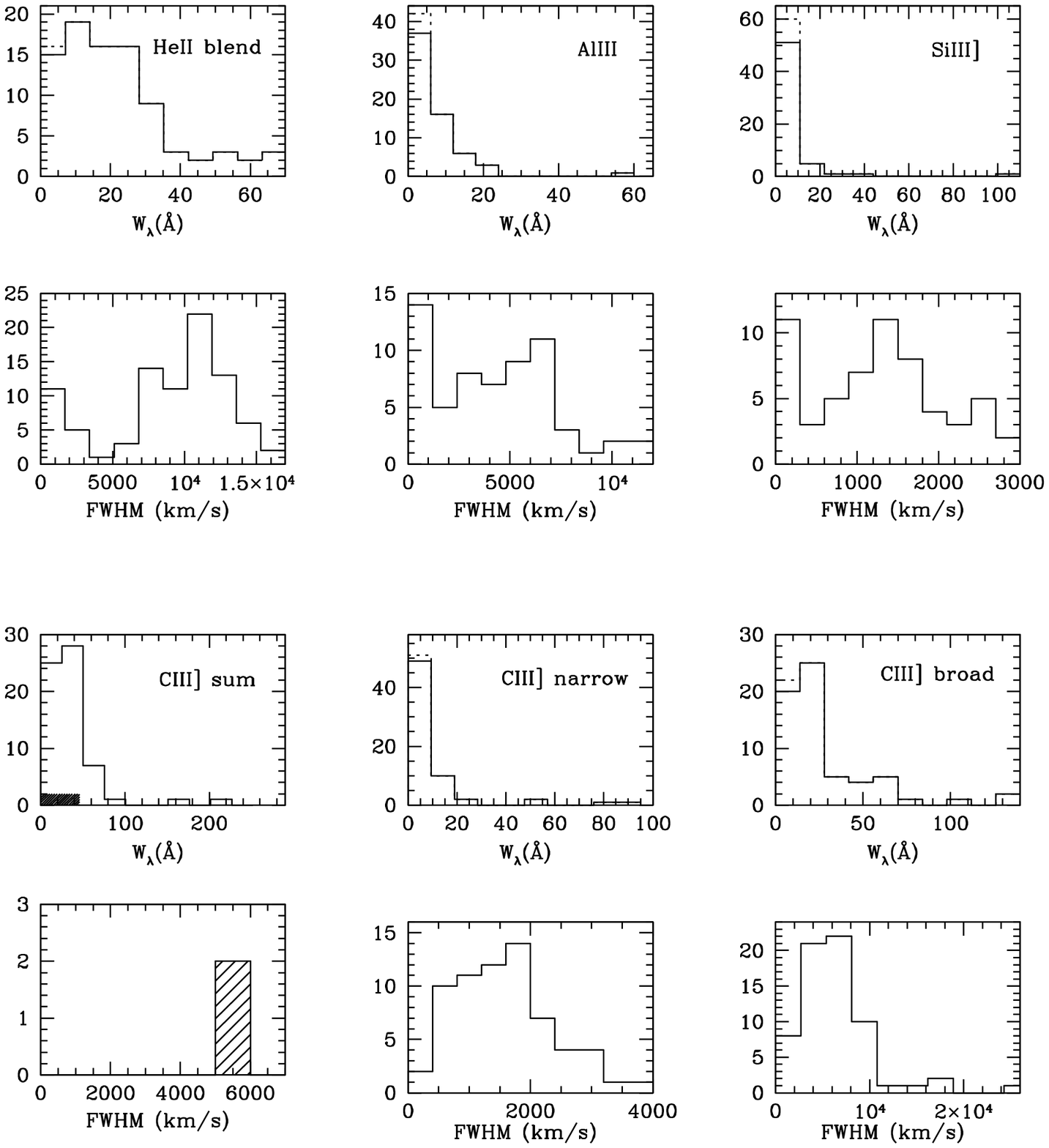}
\setcounter{figure}{1}
\caption{{\it Continued}}
\end{figure}

\clearpage

\begin{figure}
\plotone{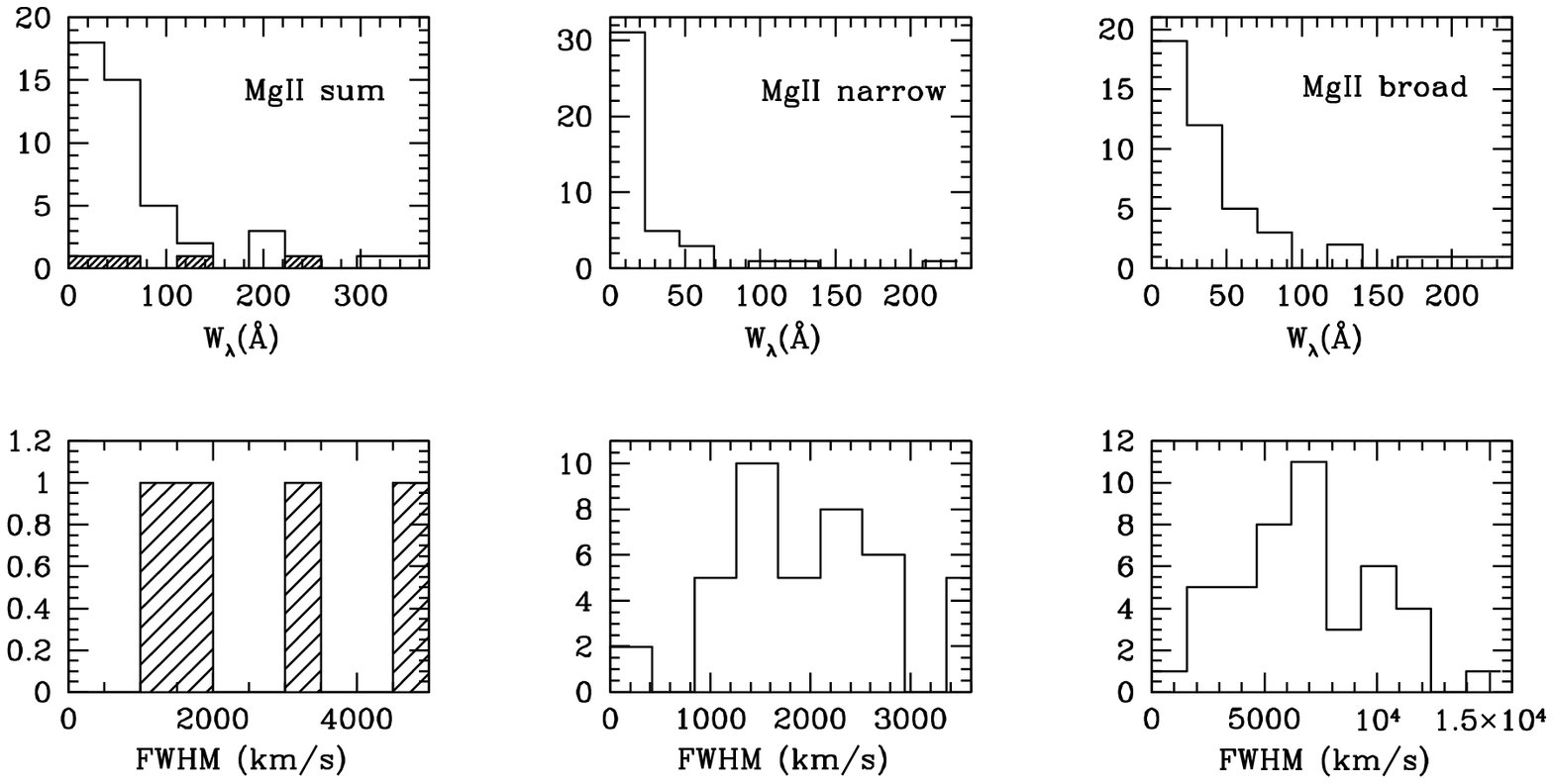}
\setcounter{figure}{1}
\caption{{\it Continued}}
\end{figure}

\clearpage

\begin{figure}
\epsscale{1.0}
\plotone{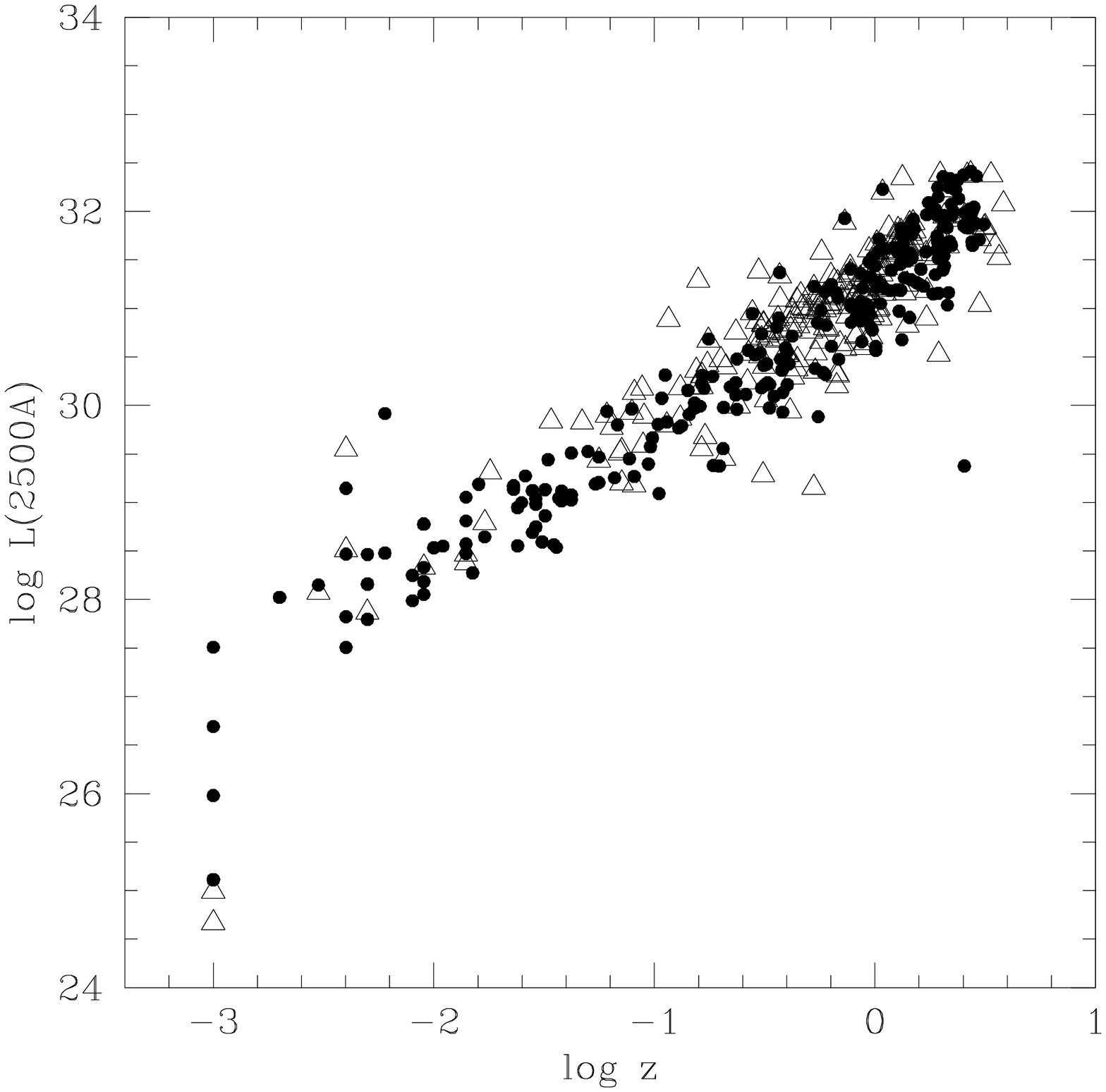}
\caption{The luminosity at 2500\AA\ versus redshift for the
post-COSTAR sample analyzed here (filled
circles) and the pre-COSTAR sample analyzed in Paper~II
(open triangles).}
\end{figure}

\clearpage

\begin{figure}
\plotone{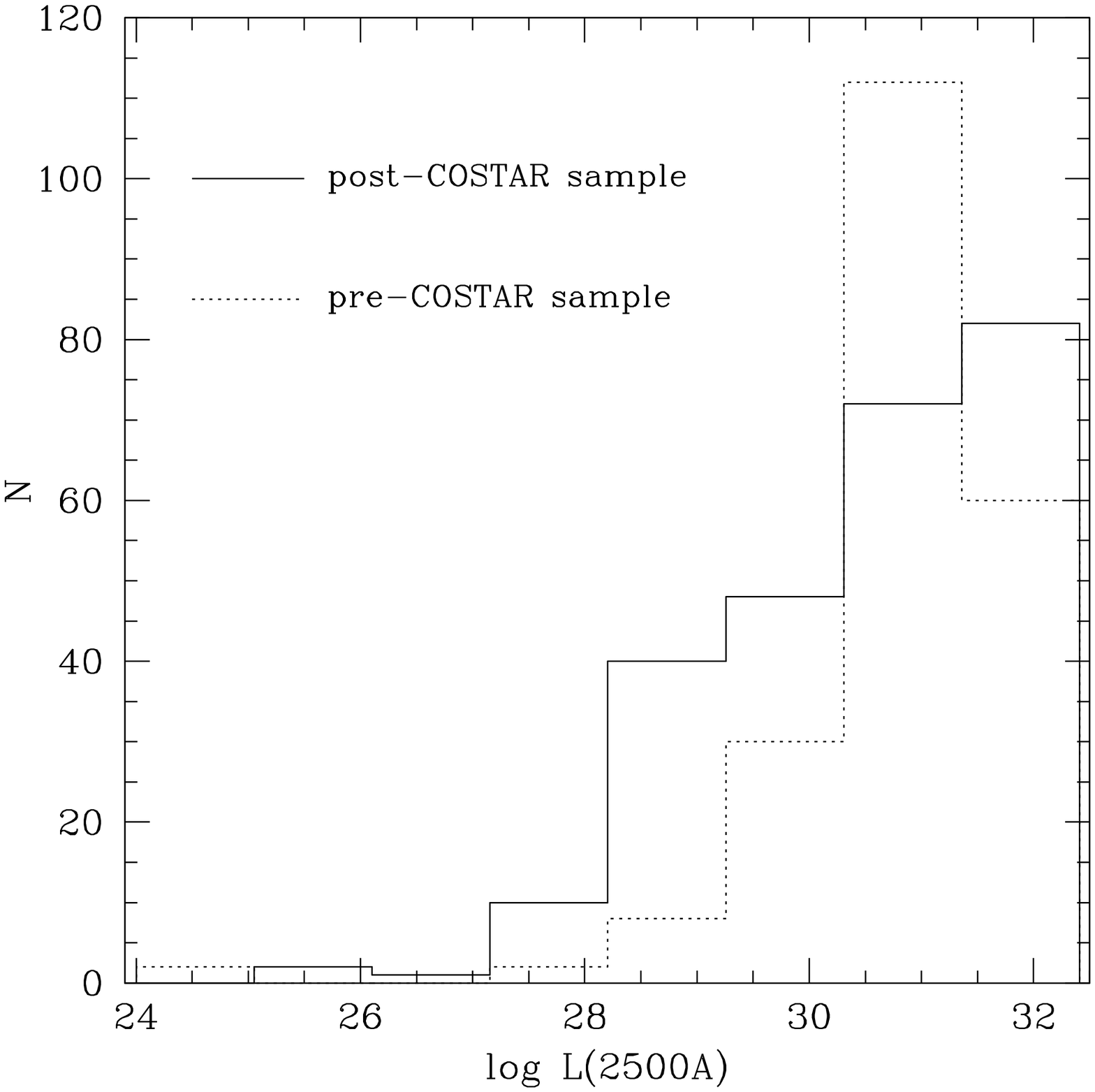}
\caption{The distribution of luminosity at 2500\AA\ for AGN from the 
post-COSTAR sample (solid line) and the pre-COSTAR sample (dotted
line). The two distributions are different at the 99.9\% level with
post-COSTAR sample having a higher percentage of low 
(log\,$L$(2500)$<30$) luminosity objects.} 
\end{figure}


\clearpage
%
%
\begin{deluxetable}{cllrrl}
\small
\tablewidth{0pt}
\tablenum{1}
\tablecaption{List of objects and spectra}
\tablehead{
\colhead{Designation(J2000)} &
\colhead{Name}&
\colhead{Type\tablenotemark{a}}&
\colhead{Redshift}&
\colhead{$N_{H}$\tablenotemark{b}}&
\colhead{Spectra}}
\startdata
0005+0203   &Q0003+0146      &Q    &0.234   &3.00   &0005+0203oa\\
0006+2012   &MRK 335         &Sy1.2&0.026   &3.70   &0006+2012oa\\
0018+1629   &QSO 0015+162    &Q    &0.553   &4.00   &0018+1629oa\\
0020+0226   &Q0017+0209      &Q    &0.401   &3.10   &0020+0226oa\\
0020+2842   &QSO 0020+287    &Q    &0.513   &4.30   &0020+2842oa\\
0029+1316   &PG 0026+129     &Q/Sy1&0.142   &4.90   &0029+1316[oa--ob]\\
0039$-$5117 &WPVS 007        &NLS1 &0.029   &2.40   &0039$-$5117oa\\
0044$-$2434 &Q0042$-$248     &Q    &0.807   &1.50   &0044$-$2434oa\\
0044+1026   &MC 0042+101     &Q    &0.583   &5.20   &0044+1026oa\\
0048+3157   &MRK 348         &Sy2  &0.015   &5.90   &0048+3157oa\\
0053+1241   &I ZW 1          &NLS1 &0.061   &5.00   &0053+1241oa\\
0103+0221   &UM 301          &Q    &0.393   &2.40   &0103+0221oa\\
0104$-$2657 &CT 336          &Q    &0.780   &1.90   &0104$-$2657oa\\
0109$-$1521 &QSO 0107$-$156  &Q    &0.861   &1.70   &0109$-$1521oa\\
0110$-$0216 &Q0107$-$0232    &Q    &0.728   &8.00   &0110$-$0216oa\\
0110$-$0219 &Q0107$-$025A    &Q    &0.960   &4.00   &0110$-$0219oA\\
            &Q0107$-$025B    &Q    &0.960   &4.00   &0110$-$0219oB\\
0120+2133   &PG 0117+213     &Q    &1.493   &4.80   &0120+2133[oa--ob]\\
0122$-$0421 &PKS 0119$-$04   &Q    &1.925   &4.20   &0122$-$0421oa\\
0124+0343   &NGC 520.48      &Q    &0.336   &2.00   &0124+0343oa\\
0139+0131   &PHL 1093        &Q    &0.260   &3.10   &0139+0131oa\\
0143+0220   &MRK 573         &Sy2  &0.017   &2.60   &0143+0220oa\\
0201$-$1132 &3C 57           &Q    &0.669   &1.80   &0201$-$1132oa\\
0206$-$0017 &MRK 1018        &Sy1.5&0.042   &2.50   &0206$-$0017oa\\
0217+1104   &PKS 0214+10     &Sy1  &0.408   &6.40   &0217+1104oa\\
0235$-$0402 &PKS 0232$-$04   &Q    &1.450   &2.20   &0235$-$0402oa\\
0238+1636*  &AO 0235+164     &BLLac&0.940   &7.60   &0238+1636ob\\ 
0241$-$0815 &NGC 1052        &Sy2  &0.005   &2.90   &0241$-$0815[oa--ob]\\
0251+4315   &S4 0248+43      &Q    &1.310  &10.10   &0251+4315oa\\
0256$-$3315 &PKS 0254$-$33   &Q    &1.915   &2.30   &0256$-$3315[oa--ob]\\
0304$-$2211 &1E 0302$-$223   &Q    &1.400   &1.80   &0304$-$2211oa\\
0320$-$1926 &0318$-$196      &Q    &0.104   &2.80   &0320$-$1926oa\\
0336+3218   &NRAO 140        &Q    &1.258  &14.20   &0336+3218oa\\
0347+0105   &IRAS 0345+0055  &Q    &0.031   &8.80   &0347+0105oa\\
0351$-$2744 &PKS 0349$-$27   &NLRG &0.066   &0.90   &0351$-$2744oa\\
0357$-$4812 &PKS 0355$-$48   &Q    &1.005   &1.10   &0357$-$4812[oa--ob]\\
0417$-$0553 &PKS 0414$-$06   &Q    &0.775   &4.30   &0417$-$0553oa\\
0423$-$0120 &PKS 0420$-$01   &Q    &0.915   &9.40   &0423$-$0120oa\\
0424+0204   &PKS 0421+01     &Q    &2.044  &10.50   &0424+0204oa\\
0452$-$2953 &IRAS 0450$-$2958&Q    &0.286   &1.30   &0452$-$2953oa\\
0456+0400   &PKS 0454+039    &Q    &1.345   &6.60   &0456+0400oa\\
0516$-$0008 &AKN 120         &Sy1  &0.033  &10.30   &0516$-$0008oa\\
0519$-$4546 &PKS 0518$-$45   &Sy1  &0.035   &4.10   &0519$-$4546oa\\
0552$-$0727 &NGC 2110    &Sy2/LINER&0.008  &18.60   &0552$-$0727oa\\
0630+6905   &HS 0624+6907    &Q    &0.370   &6.30   &0630+6905oa\\
0741+3111   &OI 363          &Q    &0.635   &4.90   &0741+3111oa\\
0742+6510   &MRK 78          &Sy2  &0.037   &4.00   &0742+6510[oa--oj]\\
0743$-$6726 &PKS 0743$-$67   &Q    &1.510  &11.90   &0743$-$6726oa\\
0804+0506   &MRK 1210        &Sy2  &0.014   &3.40   &0804+0506oa\\
0813+4813   &3C 196.0        &Q    &0.871   &4.90   &0813+4813oa\\
0830+2410   &B2 0827+24      &Q    &0.939   &3.00   &0830+2410[oa--ob]\\
0838+2453   &NGC 2622        &Sy1.8&0.029   &3.30   &0838+2453oa\\
0851+1612   &Q0848+163       &Q    &1.936   &2.40   &0851+1612oa\\
0853+5118   &NGC 2681        &Q    &0.002   &2.50   &0853+5118oa\\
\tablebreak
0909$-$0932 &QSO 0909$-$095  &Q    &0.646   &4.50   &0909$-$0932oa\\
0948+4039   &4C 40.24        &Q    &1.252   &1.30   &0948+4039oa\\
0950+3926   &PG 0947+396     &Sy1  &0.206   &1.90   &0950+3926oa\\
0955+6903   &NGC 3031        &Sy1.8&0.000   &4.30   &0955+6903[oa-oc]\\
1001+5553   &0957+561A       &Q    &1.414   &0.70   &1001+5553oA\\
            &0957+561B       &Q    &1.414   &0.70   &1001+5553oB\\
1004+0513   &PG 1001+054     &Q    &0.161   &1.80   &1004+0513oa\\
1004+2225   &PKS 1001+22     &Q    &0.974   &3.10   &1004+2225oa\\
1017$-$2046 &J03.13A         &Q    &2.545   &6.00   &1017$-$2046oA\\
            &J03.13B         &Q    &2.545   &6.00   &1017$-$2046oB\\
1019+2744   &TON 34          &Q    &1.928   &2.60   &1019+2744oa\\
1028$-$0100 &Q1026$-$0045$-$A&Q    &1.437   &4.80   &1028$-$0100oA\\
            &Q1026$-$0045$-$B&Q    &1.437   &4.80   &1028$-$0100oB\\
1030+3102   &B2 1028+313     &Q    &0.178   &0.50   &1030+3102oa\\
1034+3938   &ZW 212.025      &Sy1  &0.042   &1.40   &1034+3938oa\\
1041+0610   &4C 06.41        &Q    &1.270   &2.80   &1041+0610oa\\
1048$-$2509 &NGC 3393        &Sy2  &0.012   &5.80   &1048$-$2509oa\\
1052+6125   &4C 61.20        &Q    &0.422   &0.90   &1052+6125oa\\
1058+1951   &PKS 1055+20     &Q    &1.110   &1.80   &1058+1951oa\\
1101+1102   &MRK 728         &Sy1.9&0.036   &2.10   &1101+1102oa\\
1106+7234   &NGC 3516        &Sy1.5&0.009   &2.90   &1106+7234[oa--oe]\\
1117+4413   &PG 1114+445     &Q    &0.144   &1.90   &1117+4413[oa--ob]\\
1118+0745   &PG 1115+080A1   &Q    &1.718   &3.50   &1118+0745oA\\
            &PG 1115+080A2   &Q    &1.718   &3.50   &1118+0745oB\\
1118+4025   &PG 1115+407     &Sy1  &0.154   &1.70   &1118+4025oa\\
1119+2119   &PG 1116+215     &Q    &0.176   &1.40   &1119+2119oa\\
1121+1236   &MC 1118+12      &Q    &0.685   &2.30   &1121+1236oa\\
1124+4201   &Q1121+423       &Q    &0.234   &2.30   &1124+4201oa\\
1126+3515   &MRK 423         &Sy1.9&0.032   &1.90   &1126+3515oa\\
1127+2654   &QSO 1127+269    &Q    &0.378   &1.40   &1127+2654oa\\
1130$-$1449 &PKS 1127$-$14   &Q    &1.187   &3.80   &1130$-$1449oa\\
1135$-$0318 &Q1132$-$0302    &Q    &0.237   &3.50   &1135$-$0318oa\\
1139+3154   &NGC 3786        &Sy1.8&0.009   &2.20   &1139+3154oa\\
1141+0148   &Q1138+0204      &Q    &0.383   &2.30   &1141+0148oa\\
1147$-$0132 &Q1144$-$0115    &Q    &0.382   &2.30   &1147$-$0132oa\\
1148+1047   &1146+111B       &Q    &1.010   &3.60   &1148+1047oB\\ 
1148+1050   &1146+111C       &Q    &1.010   &3.60   &1148+1050oC\\  
1148+1046   &1146+111E       &Q    &1.100   &3.60   &1148+1046oE\\  
1148+1054   &MC 1146+111     &Q    &0.863   &3.60   &1148+1054oa\\
1151+5437   &PG 1148+549     &Q    &0.969   &0.90   &1151+5437oa\\
1151+3825   &B2 1148+387     &Q    &1.303   &2.10   &1151+3825oa\\
1153+4931   &LB 2136         &Q    &0.334   &2.10   &1153+4931oa\\
1159+2106   &TEX 1156+213    &Q    &0.349   &2.20   &1159+2106oa\\
1159+2914   &4C 29.45        &Q    &0.729   &1.50   &1159+2914oa\\
1204+2754   &GQ COMAE        &Sy1  &0.165   &1.70   &1204+2754oa\\
1204+3110   &UGC 7064        &Sy1.9&0.025   &1.60   &1204+3110oa\\
1210+3924   &NGC 4151        &Q    &0.003   &2.10   &1210+3924oa\\
1216+1748   &Q1214+1804      &Q    &0.374   &2.70   &1216+1748oa\\
1217+6407   &4C 64.15        &Q    &1.288   &2.30   &1217+6407oa\\
1219+0545   &QSO1219+057     &Q    &0.114   &1.60   &1219+0545oa\\
1221+0430   &1219+047        &Q    &0.094   &1.60   &1221+0430[oa--ob]\\
1223+1545   &1220+1601       &Q    &0.081   &2.30   &1223+1545[oa--ob]\\
1230+1223   &NGC 4486        &NLRG &0.004   &2.50   &1230+1223[ob--od]\\
1231$-$0224 &PKS 1229$-$02   &Q    &1.045   &2.30   &1231$-$0224oa\\
1237+1149   &NGC 4579        &Sy1.9&0.005   &3.00   &1237+1149oa\\
1239$-$1137 &NGC 4594        &Q    &0.004   &3.70   &1239$-$1137oa\\
1240$-$3645 &IC 3639         &Sy2  &0.011   &5.60   &1240$-$3645oa\\
1250+2631   &PG 1247+267     &Q    &2.038   &0.80   &1250+2631[oa--ob]\\
\tablebreak
1250+3016   &B2 1248+30      &Q    &1.061   &1.10   &1250+3016oa\\
1250+3125   &CSO 173         &Q    &1.020   &1.20   &1250+3125oa\\
1250+3951   &PG 1248+401     &Q    &1.030   &1.30   &1250+3951oa\\
1252+2913   &CSO 176         &Q    &0.820   &1.10   &1252+2913oa\\
1253+3105   &CSO 179         &Q    &0.780   &1.20   &1253+3105oa\\
1256$-$0547 &3C 279          &Q    &0.536   &2.20   &1256$-$0547oa\\
1256+5652   &MRK 231         &Sy1  &0.042   &1.00   &1256+5652oa\\
1301+2819   &Q1258+285       &Q    &1.360   &3.00   &1301+2819oa\\
1305$-$1033 &PKS 1302$-$102  &Q    &0.278   &3.20   &1305$-$1033oa\\
1307+0642   &3C 281          &Q    &0.602   &2.20   &1307+0642oa\\
1310+4601   &HS 1307+4617    &Q    &2.080   &1.20   &1310+4601oa\\
1312+3515   &PG 1309+355     &Sy1.2&0.184   &1.00   &1312+3515oa\\
1314+0201   &Q1311+0217      &Q    &0.306   &2.00   &1314+0201oa\\
1321+2847   &TON 156         &Q    &0.549   &1.20   &1321+2847oa\\
1323+2910   &TON 157         &Q    &0.960   &1.10   &1323+2910oa\\
1323+6541   &PG 1322+659     &Q    &0.168   &1.80   &1323+6541[oa--ob]\\
1324+0537   &IRAS 1321+0552  &Q    &0.205   &2.30   &1324+0537oa\\
1325+6515   &4C 65.15        &Q    &1.618   &1.90   &1325+6515oa\\
1331+3030   &3C 286.0        &Q    &0.849   &1.10   &1331+3030oa\\
1331+4101   &PG 1329+412     &Q    &1.930   &0.70   &1331+4101oa\\
1337+2423   &IRAS13349+2438  &Sy1  &0.108   &1.00   &1337+2423[oa--ob]\\
1338+0432   &NGC 5252        &Sy1.9&0.023   &2.00   &1338+0432[oa--ob]\\
1341+6740   &MRK 270         &Sy2  &0.009   &1.80   &1341+6740oa\\
1342$-$0053 &Q1340$-$0038    &Q    &0.326   &2.10   &1342$-$0053oa\\
1348+2622   &QSO 1348+263    &Q    &0.597   &1.10   &1348+2622oa\\
1354+3139   &B2 1351+31      &Q    &1.326   &1.20   &1354+3139oa\\
1354+1805   &PG 1352+183     &Q    &0.152   &1.80   &1354+1805oa\\
1354+0052   &PG 1352+011     &Q    &1.117   &2.00   &1354+0052oa\\
1405+2555   &PG 1402+261     &Sy1  &0.164   &1.40   &1405+2555[oa--ob]\\
1406+2223   &PG 1404+226     &Sy1  &0.098   &2.00   &1406+2223oa\\
1417+4456   &PG 1415+451     &Q    &0.114   &0.90   &1417+4456oa\\
1419$-$1310 &PG 1416$-$129   &Q    &0.129   &7.20   &1419$-$1310oa\\
1419+0628   &3C 298          &Q    &1.436   &2.00   &1419+0628oa\\
1424+2256   &QSO 1422+231A   &Q    &3.620   &2.50   &1424+2256oA\\
            &QSO 1422+231B   &Q    &3.620   &2.50   &1424+2256oB\\
            &QSO 1422+231C   &Q    &3.620   &2.50   &1424+2256oC\\
            &QSO 1422+231D   &Q    &3.620   &2.50   &1424+2256oD\\
1427+2632   &B2 1425+267     &Q    &0.366   &1.50   &1427+2632oa\\
1429+4747   &PG 1427+480     &Q    &0.221   &1.60   &1429+4747oa\\
1432$-$4410 &NGC 5643        &Sy2  &0.004   &8.50   &1432$-$4410oa\\
1437$-$0147 &Q1435$-$0134    &Q    &1.310   &3.60   &1437$-$0147[oa--ob]\\
1442+3526   &MRK 478         &NLS1 &0.079   &0.90   &1442+3526[oa--ob]\\
1442$-$1715 &NGC 5728        &Sy2  &0.009   &7.60   &1442$-$1715oa\\
1446+4035   &PG 1444+407     &Sy1  &0.267   &1.00   &1446+4035oa\\
1454$-$3747 &PKS 1451$-$375  &Q    &0.314   &6.20   &1454$-$3747oa\\
1504+6856   &B2 1503+691     &Q    &0.318   &2.20   &1504+6856oa\\
1519+2346   &LB 9612         &Q    &1.898   &3.90   &1519+2346oa\\
1519+2347   &LB 9605         &Q    &1.834   &3.90   &1519+2347oa\\
1524+0958   &PG 1522+101     &Q    &1.321   &2.60   &1524+0958oa\\
1526+4140   &NGC 5929        &Sy2  &0.008   &1.90   &1526+4140oa\\
1545+4846   &1543+489        &Q    &0.400   &1.60   &1545+4846oa\\
1559+3501   &MRK 493         &Sy1  &0.032   &2.00   &1559+3501oa\\
1617+3222   &3C 332          &Sy1? &0.152   &2.00   &1617+3222oa\\
1624+2345   &3C 336.0        &Q    &0.927   &4.50   &1624+2345oa\\
1627+5522   &PG 1626+554     &Sy1  &0.133   &1.50   &1627+5522oa\\
1629+2426   &MRK 883         &Sy1.9&0.038   &3.80   &1629+2426oa\\
1632+3737   &PG 1630+377     &Q    &1.466   &0.90   &1632+3737oa\\
1642+3948   &3C 345          &Q    &0.593   &0.80   &1642+3948oa\\
\tablebreak
1658+0515   &PKS 1656+053    &Q    &0.879   &6.10   &1658+0515oa\\
1716+5328   &PG 1715+535     &Q    &1.920   &2.40   &1716+5328oa\\
1719+4858   &ARP 102B        &LINER&0.024   &2.20   &1719+4858oa\\
1719+4804   &PG 1718+481     &Q    &1.084   &2.10   &1719+4804oa\\
1755+1820   &NGC 6500        &LINER&0.010   &6.90   &1755+1820oa\\
1800+7828   &S5 1803+78      &Q    &0.680   &3.60   &1800+7828oa\\
1842+7946   &3C 390.3        &Sy1  &0.056   &3.60   &1842+7946oa\\
1902+3159   &3C 395          &Sy1.5&0.635  &11.00   &1902+3159oa\\
1927+7358   &4C 73.18        &Q    &0.302   &7.20   &1927+7358oa\\
1959+4044   &CYGNUS A        &Q    &0.056  &33.00   &1959+4044oa\\
2118+2626   &NGC 7052        &Q    &0.014   &9.60   &2118+2626oa\\
2156+0722   &MRK 516         &Sy1.8&0.028   &4.50   &2156+0722oa\\
2242+2943   &AKN 564         &Sy1.8&0.024   &6.20   &2242+2943oa\\
2253+1608   &3C 454.3        &Q    &0.859   &6.90   &2253+1608oa\\
2254$-$1734 &MR 2251$-$178   &Sy1  &0.068   &2.70   &2254$-$1734oa\\
2303+0852   &NGC 7469        &Sy1.2&0.016   &4.80   &2303+0852oa\\
2304+0311   &PG 2302+029     &Q    &1.044   &4.90   &2304+0311oa\\

\enddata
\tablecomments{Table~1 is also available in machine-readable form in the
electronic edition of the {\it Astrophysical Journal Supplement.}}
\tablenotetext{a}{AGN type: QSO (Q), Seyfert (Sy), Narrow Line Seyfert
1 (NLS1), Narrow Line Radio Galaxy (NLRG).}
\tablenotetext{b}{$N_H$ is in units of $10^{20}$ cm$^{-2}$.}
\tablenotetext{*}{We include this BL Lac object as it shows weak
emission lines.} 
\end{deluxetable}

\clearpage
\begin{deluxetable}{clrlrr}
\tablewidth{0pt}
\tablenum{2}
\tablecaption{Representative list of objects and FOS spectra}
\tablehead{
\colhead{Spectrum} &
\colhead{Dataset}&
\colhead{Config}&
\colhead{Grating} &
\colhead{Exp. time(s)} &
\colhead{Time of obs.}
}
\startdata
0005+0203oa&  Y29C0102T&  RD&   G190H& 1379.9& Jul 21 1994\\
0006+2012oa&  Y29E0202T&  BL&   G130H& 1389.9& Dec 16 1994\\
           &  Y29E0203T&  BL&   G130H&  769.9& Dec 16 1994\\
           &  Y29E0204T&  BL&   G190H&  960.0& Dec 16 1994\\
           &  Y29E0205T&  BL&   G270H&   60.0& Dec 16 1994\\
           &  Y29E0206T&  BL&   G270H&  420.0& Dec 16 1994\\
0018+1629oa&  Y3IS0105T&  RD&   G190H& 1080.0& Jan 30 1997\\
           &  Y3IS0106T&  RD&   G190H& 2409.9& Jan 30 1997\\
           &  Y3IS0107T&  RD&   G190H& 2409.9& Jan 30 1997\\
           &  Y3IS0108T&  RD&   G190H& 2409.9& Jan 30 1997\\
           &  Y3IS0109T&  RD&   G190H& 2409.9& Jan 30 1997\\
           &  Y3IS010AT&  RD&   G190H& 2409.9& Jan 30 1997\\
           &  Y3IS010BT&  RD&   G190H& 1519.9& Jan 30 1997\\
0020+0226oa&  Y29C0202T&  RD&   G190H& 1369.9& Aug 06 1994\\
           &  Y29C0203T&  RD&   G190H&  423.9& Aug 06 1994\\
0020+2842oa&  Y3AG0102T&  RD&   G270H& 1479.9& Nov 17 1996\\
           &  Y3AG0103T&  RD&   G270H& 2430.0& Nov 17 1996\\
           &  Y3AG0104T&  RD&   G270H& 2430.0& Nov 17 1996\\
           &  Y3AG0105T&  RD&   G270H& 2430.0& Nov 17 1996\\
0029+1316oa&  Y27O0302T&  BL&   G130H&  720.0& Jul 30 1994\\
0029+1316ob&  Y2JK0102T&  BL&   G130H& 1469.9& Nov 27 1994\\
           &  Y2JK0103T&  BL&   G130H& 2149.9& Nov 27 1994\\
           &  Y2JK0104T&  BL&   G130H& 2149.9& Nov 27 1994\\
           &  Y2JK0105T&  BL&   G130H& 2149.9& Nov 27 1994\\
           &  Y2JK0106T&  BL&   G130H& 1680.0& Nov 27 1994\\
           &  Y2JK0108T&  RD&   G270H&  539.9& Nov 27 1994\\
           &  Y2JK0109T&  RD&   G190H&  827.9& Nov 27 1994\\
\enddata
\tablecomments{Table~2 is available in its entirety in the electronic
edition of the {\it Astrophysical Journal Supplement.}}
\end{deluxetable}

\clearpage

\begin{deluxetable}{lrrcr}
\small
\tablewidth{0pt}
\tablenum{3}
\tablecaption{Continuum parameters}
\tablehead{
\multicolumn{1}{c}{Designation} &
\colhead{$\Gamma_{UV}\tablenotemark{a}$}&
\colhead{Norm.\tablenotemark{b}} &
\colhead{$\lambda_{norm}$} &
\colhead{$\Gamma_{opt}\tablenotemark{a}$}\\
{\hfill (1) \hfill} & {\hfill (2) \hfill} & 
{\hfill (3) \hfill} & {\hfill (4) \hfill} &
{\hfill (5) \hfill} \\
}
\startdata
0005+0203oa &  0.17$^{+ 0.19}_{ -0.23}$ &  0.382$^{+ 0.009}_{ -0.009}$&   1804.7&...\\
0006+2012oa &  1.46$^{+ 0.02}_{ -0.01}$ &  9.579$^{+ 0.057}_{ -0.068}$&   1499.9&...\\
0018+1629oa &  1.92$^{+ 0.25}_{ -0.24}$ &  0.058$^{+ 0.001}_{ -0.001}$&   2271.3&...\\
0020+0226oa & $-0.77^{+ 0.79}_{ -0.86}$ &  0.194$^{+ 0.006}_{ -0.006}$&   2049.0&...\\
0020+2842oa &  2.08$^{+ 0.12}_{ -0.10}$ &  0.049$^{+ 0.001}_{ -0.001}$&   2227.3&...\\
0029+1316oa &  1                        &  2.905$^{+ 0.189}_{ -0.189}$&   1513.1&...\\
0029+1316ob &  1.14$^{+ 0.03}_{ -0.02}$ &  2.028$^{+ 0.016}_{ -0.018}$&   1670.2&...\\
0039$-$5117oa* &  0.28$^{+ 0.04}_{ -0.04}$ &  0.336$^{+ 0.019}_{ -0.001}$&   3000.0&  2.00$^{+ 0.05}_{  -0.02} $\\
0044+1026oa &  0.31$^{+ 0.51}_{ -0.48}$ &  0.036$^{+ 0.002}_{ -0.002}$&   2315.1&...\\
0044$-$2434oa &  0.91$^{+ 0.83}_{ -0.05}$ &  0.079$^{+ 0.006}_{ -0.000}$&   2642.7&...\\
0048+3157oa &  0.94$^{+ 0.55}_{ -0.88}$ &  0.019$^{+ 0.006}_{ -0.003}$&   2271.3&...\\
0053+1241oa &  0.89$^{+ 0.01}_{ -0.01}$ &  2.881$^{+ 0.009}_{ -0.011}$&   1798.6&...\\
0103+0221oa &  1.37$^{+ 0.44}_{ -0.46}$ &  0.267$^{+ 0.007}_{ -0.007}$&   2037.3&...\\
0104$-$2657oa &  1                        &  0.164$^{+ 0.002}_{ -0.002}$&   2273.9&...\\
0109$-$1521oa &  1.44$^{+ 0.12}_{ -0.12}$ &  0.070$^{+ 0.001}_{ -0.001}$&   2721.7&...\\
0110$-$0216oa &  1                        &  0.078$^{+ 0.001}_{ -0.001}$&   2289.6&...\\
0110$-$0219oA &  2.01$^{+ 0.08}_{ -0.32}$ &  0.100$^{+ 0.000}_{ -0.002}$&   2866.5&...\\
0110$-$0219oB &  1.86$^{+ 0.13}_{ -0.13}$ &  0.205$^{+ 0.002}_{ -0.002}$&   2866.5&...\\
0120+2133oa &  1                        &  0.565$^{+ 0.011}_{ -0.011}$&   3184.8&...\\
0120+2133ob &  1                        &  0.227$^{+ 0.020}_{ -0.020}$&   3289.1&...\\
0122$-$0421oa &  1                        &  0.086$^{+ 0.003}_{ -0.003}$&   3272.2&...\\
0124+0343oa &  0.48$^{+ 0.06}_{ -0.05}$ &  0.021$^{+ 0.000}_{ -0.000}$&   2264.5&...\\
0139+0131oa &  1.03$^{+ 0.12}_{ -0.10}$ &  0.084$^{+ 0.003}_{ -0.003}$&   1842.8&...\\
\enddata    
\tablecomments{Table~3 is available in its entirety in the electronic
edition of the {\it Astrophysical Journal Supplement.}
(*) See Notes on individual spectra in Appendix.} 
\tablenotetext{a} {The power law continuum slopes $\Gamma_{UV}$ and
$\Gamma_{opt}$ are defined as: $f_{\lambda} \propto \lambda^{-\Gamma}$.
$\Gamma_{UV}$ is fitted at $\lambda_{rest}<4200$, $\Gamma_{opt}$ at
$\lambda_{rest}>4200$. Slopes with no listed errors show the assumed slope
value in cases where only a single continuum window
was available.} 
\tablenotetext{b} {Normalization of the UV power law in units of
$10^{-14}$~erg~cm$^{-2}$~s$^{-1}$~\AA$^{-1}$, at observed wavelength
$\lambda_{norm}$.}  
\end{deluxetable}

\clearpage
\begin{deluxetable}{lrrrrc}
\tablewidth{0pt}
\tablenum{4}
\tablecaption{Representative emission line measurements}
\tablehead{
{\bf Designation} & {\bf Redshift} \\
\\
Emission Line
& {\hfill FWHM \hfill} & 
{\hfill $\Delta v_{peak}$ \hfill} & 
{\hfill $W_{\lambda}$ \hfill} 
&{\hfill Observed Flux \hfill}
& Absorption \\ 
  & {\hfill (km s$^{-1}$) \hfill} & 
{\hfill (km s$^{-1}$) \hfill} & 
{\hfill (\AA) \hfill} &
{10$^{-14}$erg\,cm$^{-2}$\,s$^{-1}$}
& Lines \\
}
\startdata
\noalign{\vskip 0.05cm}
{\bf 1106+7234oe}  &  {\bf z = 0.00880}  \\
   \\
\noalign{\vskip 0.05cm}
UV iron                   &$4250^{+   5750}_{   -250} $&$     0^{+      0}_{      0} $&$  164.40^{+     2.10}_{    -2.10}$&$  195.90 ^{+    2.50 }_{   -2.50} $&  0\\
Optical iron              & \ldots & \ldots & \ldots & \ldots & \ldots \\
Ly$\beta$                 & \ldots & \ldots & \ldots & \ldots & \ldots \\
Ly$\alpha$ narrow         &$ 950^{+     20}_{    -80} $&$   380^{+     20}_{    -20} $&$   41.60^{+     1.80}_{    -6.60}$&$   49.70 ^{+    2.10 }_{   -7.90}$&  0\\
Ly$\alpha$ broad          &$5300^{+     70}_{    -70} $&$   300^{+     40}_{    -40} $&$  225.40^{+     6.20}_{    -6.20}$&$  269.50 ^{+    7.40 }_{   -7.40} $&  2\\
\ion{N}{5}                &$3750^{+    240}_{    -90} $&$   750^{+     40}_{   -140} $&$   41.30^{+     4.80}_{    -1.90}$&$   49.40 ^{+    5.70 }_{   -2.20} $&  3\\
\ion{O}{1}                &$2800^{+    320}_{   -280} $&$  -350^{+    200}_{   -180} $&$    7.40^{+     1.90}_{    -1.60}$&$    8.90 ^{+    2.30 }_{   -1.90} $&  0\\
\ion{Si}{4} + \ion{O}{4}] &$4850^{+    140}_{   -140} $&$  -250^{+     80}_{    -80} $&$   40.20^{+     2.40}_{    -2.30}$&$   48.10 ^{+    2.90 }_{   -2.80} $&  0\\
\ion{C}{4} narrow         &$2300^{+     60}_{    -60} $&$   280^{+     30}_{    -30} $&$   63.60^{+     3.30}_{    -3.20}$&$   76.20 ^{+    4.00 }_{   -3.80} $&  0\\
\ion{C}{4} broad          &$6700^{+     50}_{    -50} $&$  -340^{+     30}_{    -30} $&$  240.80^{+     4.10}_{    -4.00}$&$  288.40 ^{+    4.90 }_{   -4.80} $&  1\\
\ion{He}{2} blend         &$6800^{+    240}_{   -300} $&$  1050^{+    180}_{   -180} $&$   27.60^{+     2.20}_{    -2.40}$&$   33.10 ^{+    2.60 }_{   -2.90} $&  0\\
\ion{Al}{3}               &$9000^{+   1700}_{  -1500} $&$  -800^{+    800}_{   -800} $&$    6.80^{+     2.20}_{    -1.80}$&$    8.10 ^{+    2.70 }_{   -2.10} $&  0\\
\ion{Si}{3}]              &$1800^{+     20}_{   -900} $&$  -440^{+     40}_{    -50} $&$   12.90^{+     0.40}_{    -7.30}$&$   15.50 ^{+    0.40 }_{   -8.70} $&  0\\
\ion{C}{3}] narrow        &$1800^{+     10}_{   -100} $&$  -740^{+     50}_{    -10} $&$   28.30^{+     0.40}_{    -3.00}$&$   33.90 ^{+    0.50 }_{   -3.60} $&  0\\
\ion{C}{3}] broad         &$7400^{+     40}_{   -260} $&$ -1950^{+     50}_{    -80} $&$   54.10^{+     0.80}_{    -4.00}$&$   64.90 ^{+    0.90 }_{   -4.80} $&  0\\
\ion{Mg}{2} narrow        &$2850^{+     50}_{    -50} $&$  -620^{+     30}_{    -30} $&$   51.90^{+     1.70}_{    -1.70}$&$   62.40 ^{+    2.10 }_{   -2.10} $&  0\\
\ion{Mg}{2} broad         &$7550^{+     30}_{   -260} $&$  -650^{+     20}_{   -100} $&$  103.40^{+     1.10}_{    -5.60}$&$  124.40 ^{+    1.30 }_{   -6.80} $&  2\\

\enddata
\tablecomments{Table~4 is available in its entirety in the electronic
edition of the {\it Astrophysical Journal Supplement.} A portion is
shown here for guidance regarding its form and contents. We present
here line measurements for one, example spectrum 1106+7234oe. All
measurements are rest frame except for flux. Spectral modeling of this
object is shown in Fig.1} 
\end{deluxetable}


\clearpage
\voffset=-2.5cm
\hoffset=-1.5cm

\begin{deluxetable}{lrrrrrrrrrrr}
\small

\tablewidth{0pt}
\tablenum{5}
\tablecaption{Rest frame emission line parameter distributions}
\tablehead{
\multicolumn{3}{c}{}& \multicolumn{5}{c}{$W_{\lambda}$ (\AA)} &
\multicolumn{4}{c}{FWHM (km~s$^{-1}$)} \\
\cline{4-8} \cline{10-12} \\
\multicolumn{3}{c}{}& \multicolumn{3}{c}{Detected} & 
\multicolumn{2}{c}{Kaplan-Meier} &
\multicolumn{4}{c}{Detected} \\
Emission Line & Total & Limits &Mean & {\hfill SD \hfill}  &
Median & 
{\hfill Mean \hfill} & Median &
Num & 
{\hfill Mean \hfill} & {\hfill SD \hfill} &{\hfill Median \hfill}\\
{\hfill (1) \hfill} & {\hfill (2) \hfill} & 
{\hfill (3) \hfill} & {\hfill (4) \hfill} &
{\hfill (5) \hfill} & {\hfill (6) \hfill} & 
{\hfill (7) \hfill} & {\hfill (8) \hfill} &
{\hfill (9) \hfill} & {\hfill (10) \hfill} & 
{\hfill (11) \hfill}& {\hfill (12) \hfill}}
\startdata
UV iron                                           &  77&  15&   67$\pm$  11&   87&  54&  54$\pm$ 7&  45& 63 &   4537 $\pm$    716  &    5683 & 3500\\
Optical iron                                      &  18&   2&  103$\pm$  42&  169&  18&  91$\pm$29&  24& 16 &   5809 $\pm$    1860 &    7438 & 7200\\
Lyman $\beta$ + \ion{O}{6} $\lambda$1035          &  77&   4&   22$\pm$   7&   57&  10&  21$\pm$ 6&  10&  73&  7110$\pm$  1068&  9129&5800\\
{Lyman $\alpha$\,$\lambda$1216 \hfill single}\tablenotemark{a}&  13&   0&  710$\pm$ 360& 1299& 145&    ...    & ...& 13 &   4304 $\pm$    1539 &    5549 & 4100\\
{\hfill narrow\tablenotemark{c}}                  &  96&   1&   31$\pm$   5&   51&  19&  39$\pm$4 &  19& 96 &   2526 $\pm$     284 &    2787 & 2365\\
{\hfill broad}                                    &  97&   0&
83$\pm$  13&  126&  64&    ...  & ...& 96 &  10993 $\pm$    1242 &   12167 &11600\\
{\hfill sum}\tablenotemark{b,d}                   & 108&   0&  153$\pm$  28&  290&  87&    ...  & ...& ...&  ...          &  ... &...\\
\ion{N}{5} $\lambda$1241.5                        & 108&   8&   12$\pm$   2&   19&   7&  12$\pm$ 1&   6&  98 &   4038 $\pm$     481 &    4762 & 3475\\
\ion{O}{1} $\lambda$1305                          & 102&   8&    4$\pm$   1&    6&   2&   3$\pm$ 1&   2&  92 &   2722 $\pm$     343 &    3294 & 2450\\
\ion{Si}{4} + \ion{O}{4}] $\lambda$1400           & 102&   0&   12$\pm$   2&   17&   9&    ...  & ...&  102 &   5329 $\pm$     594 &    5997 & 5400\\
{\ion{C}{4}\,$\lambda$1549 \hfill single}\tablenotemark{a}&   4&   0&   42$\pm$  27&   55&  32&    ...  & ...&   4 &   4525 $\pm$    2683 &    5366 & 5100\\
{\hfill narrow}                                   &  92&   2&   30$\pm$   5&   45&  22&  30$\pm$ 3&  22&  89 &   2670 $\pm$     316 &    2979 & 2600\\
{\hfill broad}                                    &  92&   2&   63$\pm$  10&   90&  43&  62$\pm$ 7&  42&  90 &  10281 $\pm$    1203 &   11413 & 9910\\
{\hfill sum}\tablenotemark{b}                     &  94&   0&   91$\pm$  13&  125&  67&     ... & ...&  ..&. ...          & ...&...\\
\ion{He}{2} $\lambda$1640                         &  89&   1&   22$\pm$   3&   27&  18&  22$\pm$ 2&  19&  88 &   9037 $\pm$    1071 &   10050 &10000\\
\ion{Al}{3}  $\lambda$1859                        &  68&   5&    7$\pm$   1&   11&   5&   7$\pm$ 1&   4&  62 &   4341 $\pm$     673 & 5307 & 4600\\
\ion{Si}{3}]$\lambda$1892                         &  59&   9&    6$\pm$   2&   17&   2&   6$\pm$ 2&   2&  59 &   1306 $\pm$     201 &    1547 & 1350\\
{\ion{C}{3}]\,$\lambda$1909 \hfill single}\tablenotemark{a}& ...& ...&      ...     &  ...& ...&    ...  & ...& ...&   ...         &   ...&...\\
{\hfill narrow}                                   &  68&   2&   13$\pm$   3&   28&   5&  12$\pm$ 3&   5&  66 &   1614 $\pm$     222 &    1807 & 1575\\
{\hfill broad\tablenotemark{e}}                   &  68&   2&   27$\pm$   5&   36&  20&  26$\pm$ 3&  20&  66 &   6547 $\pm$     965 &    7840 & 5825\\
{\hfill sum}\tablenotemark{b,f}                   &  66&   0&   35$\pm$   6&   48&  27&    ...  & ...& ...&   ...         &   ...&...\\
{\ion{Mg}{2}\,$\lambda$2800 \hfill single}\tablenotemark{a}&   4&   0&  104$\pm$  76&  152&  90&    ...  & ...&   4&  2887$\pm$ 1846&  3692&2525\\
{\hfill narrow}                                   &  44&   0&   37$\pm$  13&   86&  17&    ...  & ...&  42 &   2050 $\pm$     347 &    2252 & 2050\\
{\hfill broad}                                    &  44&   0&   47$\pm$  11&   70&  27&    ...  & ...&  44 &   6751 $\pm$    1136 &    7535 & 6550\\
{\hfill sum}\tablenotemark{b}                     &  47&   0&   86$\pm$  21&  143&  50&    ...  & ...& ...&   ...         &   ...&...\\
$[$\ion{Ne}{5}] $\lambda$3426                     &  21&   1&   14$\pm$   7&   32&   3&  13$\pm$ 6&   3&  18&   940$\pm$   306&  1298&700\\
$[$\ion{O}{2}] $\lambda$3728                      &  21&   0&  138$\pm$ 102&  469&   4&    ...  & ...&  20&  1008$\pm$   343&  1534&570\\
$[$\ion{Ne}{3}] $\lambda$3869                     &  21&   0&   33$\pm$  19&   87&   6&    ...  & ...&  21&  1318$\pm$   440&  2017&800\\
H$\delta$  $\lambda$4101.7                        &  17&   2&   42$\pm$  22&   85&   6& 37$\pm$ 16&   4& 14 &   2654 $\pm$    1177 &    4404 & 1350\\
$[$S\,II] $\lambda$4072.5                         &  17&  10&   38$\pm$   9&   23&  27& 16$\pm$10&   1& 7 &    717 $\pm$     384 &    1017 &  600\\
H$\gamma$ $\lambda$4340.5                         &  15&   1&   21$\pm$  10&   37&  11&  20$\pm$ 7  &  10& 14 &   1309 $\pm$     405 &    1514 & 1325\\
$[$\ion{O}{3}] $\lambda$4363                      &  15&   3&   13$\pm$   8&   27&   6&  11$\pm$ 5  &   1& 12 &    663 $\pm$     240 &     831 &  500\\
\ion{He}{2} $\lambda$4686                         &  17&   2&   10$\pm$   5&   18&   3&  9$\pm$ 3  &   2&  15 &   2681 $\pm$    1230 &    4764 & 1000\\
{H$\beta$\,$\lambda$4861 \hfill single}\tablenotemark{a} &   5&   0&   31$\pm$  22&   48&  13&    ...  & ...&  5 &   3220 $\pm$    2190 &    4897 & 2100\\
{\hfill narrow}                                   &  12&   0&   29$\pm$  12&   42&  20&    ...  & ...&   12 &   1132 $\pm$     411 &    1423 & 1010\\
{\hfill broad}                                    &  12&   0&   90$\pm$  37&  129&  50&    ...  & ...&   12 &   9090 $\pm$    3491 &   12094 & 5150\\
{\hfill sum}\tablenotemark{b}                     &  17&   0&   94$\pm$  34&  141&  59&    ...  & ...& ...&    ...     ...&  ...&...\\
$[$\ion{O}{3}] $\lambda$4959                      &  17&   1&   35$\pm$  17&   69&  10& 33$\pm$13 &   7&  16 &   1045 $\pm$     325 &    1300 &  890\\
$[$\ion{O}{3}] $\lambda$5007                      &  16&   0&  105$\pm$  52&  209&  31&  ...  & ...&   16 &   1021 $\pm$     297 &    1190 & 1015\\
\ion{He}{1}  $\lambda$5875.6                      &  18&   2&   15$\pm$   5&   18&  13&  13$\pm$ 2&  11& 16 &   2254 $\pm$     684 &    2737 & 2400\\
$[$N\,II]$\lambda$6548                            &  26&   1&   25$\pm$   7&   37&  13&  24$\pm$ 5&   7&  24 &    615 $\pm$     161 &     786 &  330\\
{H$\alpha$\,$\lambda$6563  \hfill single}\tablenotemark{a}& ...& ...&     ...      &  ...& ...&   ...   &... & ...&    ...     ...&  ...&...\\
{\hfill narrow}                                   &  26&   0&  120$\pm$  34&  172&  82&    ...  & ...&  26 &   1362 $\pm$     335 &    1708 &  950\\
{\hfill broad}                                    &  26&   0&  190$\pm$  53&  265& 157& 183$\pm$35& 145&  25 &   4918 $\pm$    1289 &    6446 & 3400\\
{\hfill sum}\tablenotemark{b}                     &  26&   0&  203$\pm$  79&  404& 242&    ...  & ...& ...&   ...      ...&  ...&...\\
$[$N\,II]$\lambda$6583                            &  26&   2&   43$\pm$  14&   69&  28&  40$\pm$10&  23&  24 &   518  $\pm$     131 &     642 &  350\\
$[$S\,II]$\lambda$6716.4                          &  21&   1&   17$\pm$   5&   23&  14&  16$\pm$3  &  13&  20 &   603  $\pm$     193 &     864 &  475\\
$[$S\,II]$\lambda$6731                            &  20&   1&   16$\pm$   6&   26&   8&  15$\pm$4  &   7&  20 &   376  $\pm$      98 &     438 &  350\\
\enddata
\tablenotetext{a}{The distribution for single Gaussian component models 
are tabulated separately from narrow and broad components.
$^b$ The distribution  of the sum of the broad and
narrow component \ew\ measurements included with the single component
measurements.
$^c$Means, medians, and SD of \ew\  calculated excluding 
objects with high equivalent width measurements (PKS~0518-45 
and NGC~3031).
$^d$Means, medians, and SD of \ew\ calculated excluding 
objects  with high equivalent width measurements  
(NGC~5728 and NGC~5643).
$^e$Means, medians, and SD of \ew\ calculated excluding 
objects with high equivalent width measurements (NGC~5252, NGC~4579,
 and NGC 5929).
$^f$Means, medians, and SD of \ew\ calculated excluding 
objects  with high equivalent width measurements (NGC~5252, NGC~4579,
and NGC~5728).}
\tablenotetext{}{(1) Emission line or line blend, (2) total number of
emission lines modeled, (3) number of upper limits, 
(4) mean \ew\  of detected emission lines, (5) standard
deviation (SD) of \ew\  measurements for detected emission lines, (6)
median of \ew\ for detections,  (7)-(8) Kaplan-Meier
reconstructed mean and median of \ew\ distribution when upper limits
are present.  (9)--(12) The number, mean and median of
the distribution of FWHM of the Gaussian components used to model each
emission feature. 
}

\end{deluxetable}



\end{document}